\documentclass[%
reprint,
superscriptaddress,
 amsmath,amssymb,
 aps,
altaffilletter,
]{revtex4-2}

\usepackage{graphicx}
\usepackage{dcolumn}
\usepackage{bm}
\usepackage{hyperref}

\usepackage[caption=false]{subfig}

\usepackage{longtable}
\usepackage{isotope}
\usepackage{supertabular}
\usepackage{tablefootnote}
\usepackage{threeparttable}
\usepackage[referable]{threeparttablex}
\usepackage{placeins}
\usepackage{color,soul}
\usepackage{floatrow}
\usepackage{afterpage}
\usepackage{filecontents}

\begin{document}


\title{Decay Spectroscopy of $^{160}$Eu: Quasiparticle Configurations of Excited States and Structure of $K^\pi$=$4^+$ Band-heads in $^{160}$Gd}

\author{D. Yates}
\email{dyates@triumf.ca.}
\affiliation{TRIUMF, Vancouver, British Columbia V6T 2A3, Canada}
\affiliation{Department of Physics and Astronomy, University of British Columbia, Vancouver, British Columbia V6T 1Z4, Canada}

\author{R. Kr\"ucken}

\affiliation{TRIUMF, Vancouver, British Columbia V6T 2A3, Canada}
\affiliation{Department of Physics and Astronomy, University of British Columbia, Vancouver, British Columbia V6T 1Z4, Canada}
\affiliation{Lawrence Berkeley National Laboratory, Berkeley, California 94720, U.S.A.}

\author{I. Dillmann}
\affiliation{TRIUMF, Vancouver, British Columbia V6T 2A3, Canada}
\affiliation{Department of Physics and Astronomy, University of Victoria, Victoria, British Columbia V8P 5C2, Canada}

\author{P.E. Garrett}
\affiliation{Department of Physics, University of Guelph, Guelph, Ontario N1G 2W1, Canada}

\author{B. Olaizola}
\altaffiliation{Present address: Instituto de Estructura de la Materia, CSIC, Serrano 113-bis, E-28006, 
Madrid, Spain.}
\affiliation{TRIUMF, Vancouver, British Columbia V6T 2A3, Canada}

\author{V. Vedia}
\affiliation{TRIUMF, Vancouver, British Columbia V6T 2A3, Canada}

\author{F.A. Ali}
\affiliation{Department of Physics, University of Guelph, Guelph, Ontario N1G 2W1, Canada}
\affiliation{Department of Physics, College of Education, University of Sulaimani, Kurdistan Region, Iraq}

\author{C. Andreoiu}
\affiliation{TRIUMF, Vancouver, British Columbia V6T 2A3, Canada}
\affiliation{Department of Chemistry, Simon Fraser University, Burnaby, British Columbia V5A 1S6, Canada}

\author{W. Ashfield}
\altaffiliation{Present address: Department of Physics, Montana State University, Bozeman, Montana 59717, U.S.A.}
\affiliation{Department of Physics, Reed College, Portland, Oregon 97202, U.S.A.}

\author{G.C. Ball}
\affiliation{TRIUMF, Vancouver, British Columbia V6T 2A3, Canada}

\author{Z. Beadle}
\affiliation{Department of Physics, Reed College, Portland, Oregon 97202, U.S.A.}

\author{N. Bernier}
\altaffiliation{Present address: Department of Physics and Astronomy, University of the Western Cape, Bellville-7535, South Africa.}
\affiliation{TRIUMF, Vancouver, British Columbia V6T 2A3, Canada}
\affiliation{Department of Physics and Astronomy, University of British Columbia, Vancouver, British Columbia V6T 1Z4, Canada}

\author{S.S. Bhattacharjee}
\affiliation{TRIUMF, Vancouver, British Columbia V6T 2A3, Canada}

\author{H. Bidaman}
\affiliation{Department of Physics, University of Guelph, Guelph, Ontario N1G 2W1, Canada}

\author{\mbox{V. Bildstein}}
\affiliation{Department of Physics, University of Guelph, Guelph, Ontario N1G 2W1, Canada}

\author{D. Bishop}
\affiliation{TRIUMF, Vancouver, British Columbia V6T 2A3, Canada}

\author{M. Bowry}
\altaffiliation{Present address: School of Computing, Engineering, and Physical Sciences, University of the West of Scotland, Paisley PA1 2BE, Scotland, U.K.}
\affiliation{TRIUMF, Vancouver, British Columbia V6T 2A3, Canada}

\author{C. Burbadge}
\altaffiliation{Deceased.}
\affiliation{Department of Physics, University of Guelph, Guelph, Ontario N1G 2W1, Canada}

\author{R. Caballero-Folch}
\affiliation{TRIUMF, Vancouver, British Columbia V6T 2A3, Canada}

\author{D.Z. Chaney}
\altaffiliation{Department of Physics and Astronomy, University of Tennessee, Knoxville, Tennessee 37996, U.S.A.}
\affiliation{Department of Physics, Tennessee Technological University, Cookeville, Tennessee 38505, U.S.A.}

\author{D.C. Cross}
\affiliation{Department of Chemistry, Simon Fraser University, Burnaby, British Columbia V5A 1S6, Canada}

\author{A. Diaz Varela}
\affiliation{Department of Physics, University of Guelph, Guelph, Ontario N1G 2W1, Canada}

\author{M.R. Dunlop}
\affiliation{Department of Physics, University of Guelph, Guelph, Ontario N1G 2W1, Canada}

\author{\mbox{R. Dunlop}}
\affiliation{Department of Physics, University of Guelph, Guelph, Ontario N1G 2W1, Canada}

\author{\mbox{L.J. Evitts}}
\altaffiliation{Present Address: Nuclear Futures Institute, Bangor University, Bangor, Gwynedd, LL57 2DG, U.K.}
\affiliation{TRIUMF, Vancouver, British Columbia V6T 2A3, Canada}
\affiliation{Department of Physics, University of Surrey, Guildford GU2 7XH, U.K.}

\author{F.H. Garcia}
\altaffiliation{Present Address: Lawrence Berkeley National Laboratory, Berkeley, California 94720, U.S.A.}
\affiliation{Department of Chemistry, Simon Fraser University, Burnaby, British Columbia V5A 1S6, Canada}

\author{A.B. Garnsworthy}
\affiliation{TRIUMF, Vancouver, British Columbia V6T 2A3, Canada}

\author{S. Georges}
\affiliation{TRIUMF, Vancouver, British Columbia V6T 2A3, Canada}

\author{S.A. Gillespie}
\altaffiliation{Present address: FRIB, Michigan State University, East Lansing, Michigan 48824, U.S.A.}
\affiliation{TRIUMF, Vancouver, British Columbia V6T 2A3, Canada}

\author{G. Hackman}
\affiliation{TRIUMF, Vancouver, British Columbia V6T 2A3, Canada}

\author{\mbox{J. Henderson}}
\altaffiliation{Present address: Department of Physics, University of Surrey, Guildford, Surrey GU2 7XH, U.K.}
\affiliation{TRIUMF, Vancouver, British Columbia V6T 2A3, Canada}

\author{S. Jigmeddorj}
\affiliation{Department of Physics, University of Guelph, Guelph, Ontario N1G 2W1, Canada}

\author{J. Lassen}
\affiliation{TRIUMF, Vancouver, British Columbia V6T 2A3, Canada}
\affiliation{Department of Physics and Astronomy, University of Manitoba, Winnipeg, Manitoba R3T 2N2, Canada}

\author{R.~Li}
\affiliation{TRIUMF, Vancouver, British Columbia V6T 2A3, Canada}

\author{B.K. Luna}
\altaffiliation{Department of Physics and Astronomy, University of Tennessee, Knoxville, Tennessee 37996, U.S.A.}
\affiliation{Department of Physics, Tennessee Technological University, Cookeville, Tennessee 38505, U.S.A.}

\author{A.D. MacLean}
\affiliation{Department of Physics, University of Guelph, Guelph, Ontario N1G 2W1, Canada}

\author{C.R. Natzke}
\affiliation{TRIUMF, Vancouver, British Columbia V6T 2A3, Canada}
\affiliation{Department of Physics, Colorado School of Mines, Golden, Colorado 80401, U.S.A.}

\author{C.M. Petrache}
\affiliation{University Paris-Saclay, CNRS/IN2P3, IJCLab, 91405 Orsay, France}

\author{A.J. Radich}
\affiliation{Department of Physics, University of Guelph, Guelph, Ontario N1G 2W1, Canada}

\author{M.M. Rajabali}
\affiliation{Department of Physics, Tennessee Technological University, Cookeville, Tennessee 38505, U.S.A.}

\author{P.H. Regan}
\affiliation{Department of Physics, University of Surrey, Guildford, Surrey GU2 7XH, U.K.}
\affiliation{Marine, Medical and Nuclear Department, National Physical Laboratory, Teddington TW11 0LW, U.K.}

\author{Y. Saito}
\affiliation{TRIUMF, Vancouver, British Columbia V6T 2A3, Canada}
\affiliation{Department of Physics and Astronomy, University of British Columbia, Vancouver, British Columbia V6T 1Z4, Canada}

\author{J. Smallcombe}
\altaffiliation{Present address: Advanced Science Research Center, Japan Atomic Energy Agency (JAEA), Tokai, Ibaraki 319-1995, Japan.}
\affiliation{TRIUMF, Vancouver, British Columbia V6T 2A3, Canada}

\author{J.K. Smith}
\altaffiliation{Present address: Department of Physics and Astronomy, University of Notre Dame, Notre Dame, Indiana 46656, U.S.A.}
\affiliation{Department of Physics, Reed College, Portland, Oregon 97202, U.S.A.}

\author{M. Spieker}
\altaffiliation{Present address: Department of Physics, Florida State University, Tallahassee, Florida 32306, U.S.A.}
\affiliation{Institut f\"ur Kernphysik, Universit\"at zu K\"oln, 50937 K\"oln, Germany}

\author{C.E. Svensson}
\affiliation{Department of Physics, University of Guelph, Guelph, Ontario N1G 2W1, Canada}

\author{A.~Teigelh\"ofer}
\affiliation{TRIUMF, Vancouver, British Columbia V6T 2A3, Canada}
\affiliation{Department of Physics and Astronomy, University of Manitoba, Winnipeg, Manitoba R3T 2N2, Canada}

\author{K. Whitmore}
\affiliation{Department of Chemistry, Simon Fraser University, Burnaby, British Columbia V5A 1S6, Canada}

\author{T. Zidar}
\affiliation{Department of Physics, University of Guelph, Guelph, Ontario N1G 2W1, Canada}

\date{\today}


\begin{abstract}
\noindent \textbf{Background:} Detailed spectroscopy of neutron-rich, heavy, deformed nuclei is of broad interest for nuclear astrophysics and nuclear structure. Nuclei in the r-process path and following freeze-out region impact the resulting r-process abundance distribution, and the structure of nuclei midshell in both proton and neutron number helps to understand the evolution of subshell gaps and large deformation in these nuclei.

\noindent \textbf{Purpose:} To improve the understanding of the nuclear structure of $^{160}$Gd, specifically the $K^\pi$=$4^+$ bands, as well as study the $\beta$-decay of $^{160}$Eu into $^{160}$Gd.

\noindent \textbf{Methods:} High-statistics decay spectroscopy of $^{160}$Gd resulting from the $\beta$-decay of $^{160}$Eu was collected using the GRIFFIN spectrometer at the TRIUMF-ISAC facility. 

\noindent \textbf{Results:} Two new excited states and ten new transitions were observed in $^{160}$Gd. The $\beta$-decaying half-lives of the low- and high-spin isomer in $^{160}$Eu were determined, and the low-spin state's half-life was measured to be $t_{1/2}=26.0(8)$~s, $\sim$16\% shorter than previous measurements. Lifetimes of the two $K^\pi$=$4^+$ band-heads in $^{160}$Gd were measured for the first time, as well as $\gamma$-$\gamma$ angular correlations and mixing ratios of intense transitions out of those band-heads.

\noindent \textbf{Conclusions:} Lifetimes and mixing ratios suggest that the hexadecapole phonon model of the $K^\pi$=$4^+$ band-heads in $^{160}$Gd is preferred over a simple two-state strong mixing scenario, although further theoretical calculations are needed to fully understand these states. Additionally, the 1999.0 keV state in $^{160}$Gd heavily populated in $\beta$-decay is shown to have positive parity, which raises questions regarding the structure of the high-spin $\beta$-decaying state in $^{160}$Eu. 

\end{abstract}

\maketitle



\section{Introduction}

Studies of deformed midshell nuclei around $A=160$ are of particular interest for nuclear astrophysics and nuclear structure. 
The reaction flow of the rapid neutron capture process ($r$-process) proceeds far from stability. However, nuclei close to stability are involved in the ``freeze-out" path that follows the $r$-process when the nuclei decay back towards stability and can provide insight into the formation of the ``rare-earth peak" around $A=165$ resulting from the deformed shell gaps around 
$N=98$. 
A similar but less investigated abundance peak is formed around $A=100$ with deformed nuclei around $N=70$.

The nuclear structure of nuclei near both the $N=98$ and $Z=60$ deformed shell gaps is also of significant interest. The study of the isotopic and isotonic evolution of the deformation in this region may shed light on the subshell structure in these heavy, deformed nuclei. The even-$N$ neutron-rich gadolinium isotopes ($Z=64$) are some of the most quadrupole-deformed nuclei in this region, with $^{160}$Gd being the most deformed at $\beta_2=0.35109(44)$ \cite{PRITYCHENKO_adndt_2016}.

The structure of $^{160}$Gd ($N=96$), while studied extensively via scattering and transfer reactions (\cite{Govor}, for example) thanks to its proximity to stability, has until recently not been thoroughly studied via $\beta$-decay population from $^{160}$Eu. A recent study \cite{Hartley_PRL,Hartley_PRC} has significantly expanded  and improved on the conflicting level scheme from nearly 40 years prior \cite{Dauria,Morcos} and discovered that there are two $\beta$-decaying states in $^{160}$Eu. Their analysis assigned the ground-state as spin ($5^-$) and the isomeric-state at 93.0(12)~keV as spin ($1^-$).

Like the lighter even-even Gd nuclei $^{154-158}$Gd, $^{160}$Gd exhibits two low-lying $K^\pi$=$4^+$ bands  which have been considered as hexadecapole phonons \cite{Burke}. The $K^\pi$=$4^+$ bands in $^{160}$Gd have band-heads at 1070.8 keV and 1483.4 keV, and are strongly populated in the $\beta$-decay of the $J^\pi$=$(5)$ $\beta$-decaying state of $^{160}$Eu. The lower $K^\pi$=$4^+$ band was further studied by Hartley \textit{et al.} \cite{Hartley_gd_coulex} via unsafe Coulomb excitation, extending the rotational band built on the 1070.8 keV $4^+$ state to $I^\pi = 18^+$. 
In that study, an unusual alignment behavior was identified and interpreted as resulting from a quenching of static neutron pairing with a predominant two-quasi-neutron configuration. This observation is in tension with the apparent two-neutron and two-proton quasiparticle configurations present in the $4^+$ band-heads in $^{160}$Gd, and further studies of the underlying quasiparticle structure and potential band mixing are warranted.

Here we report on the observation of $\gamma$-rays in $^{160}$Gd following the $\beta$-decay of $^{160}$Eu.
Newly observed transitions and lifetimes of the two $K^\pi$=$4^+$ states are reported, as well as $\gamma$-$\gamma$ angular correlations for some of their depopulating transitions which provide critical new insights into the structure of and interplay between the $K^\pi$=$4^+$ bands in $^{160}$Gd. Lifetime and mixing ratio measurements relating to the 1999.0 keV state in $^{160}$Gd also provide insight into the structure of the parent $^{160}$Eu.


\section{Experiment}\label{Sec:Exp}
The present work is part of an experimental campaign using $\beta$-decay of $^{160-166}$Eu to study the nuclear structure of $^{160-166}$Gd using the GRIFFIN spectrometer \cite{GRIFFIN_NIM,GRIFFIN_DAQ} at the TRIUMF-ISAC facility \cite{ISAC_overview}. The cyclotron delivered a 9.8 $\mu$A beam of 480 MeV protons onto a UC$_x$ target at the ISAC target station. Products of the spallation reactions diffused to the surface of the UC$_x$ target where they are ionized and transported to experimental areas.

Crucial to the production of clean neutron-rich lanthanide beams is the ion guide laser ion source (IG-LIS) \cite{IGLIS}. The electrostatic repeller electrode between the target transfer line and laser ionization region used by IG-LIS allows for rejection of surface-ionized species from the target that would otherwise contaminate and overwhelm the ions of interest in the beam. 
Neutral atoms, including those of europium, proceed past the repeller electrode and are then ionized via an element-selective two-step resonant laser ionization. 
The isotope of interest is selected via a high-resolution mass separator ($M/\delta M\approx2000$) and delivered to the GRIFFIN facility.
Without this suppression of contaminating ions and the selective laser ionization with the IG-LIS, decay spectroscopy of neutron-rich europiums would be seriously hampered at an ISOL-type facility.

\begin{figure}[!htb]
    \includegraphics[width=1.0\columnwidth]{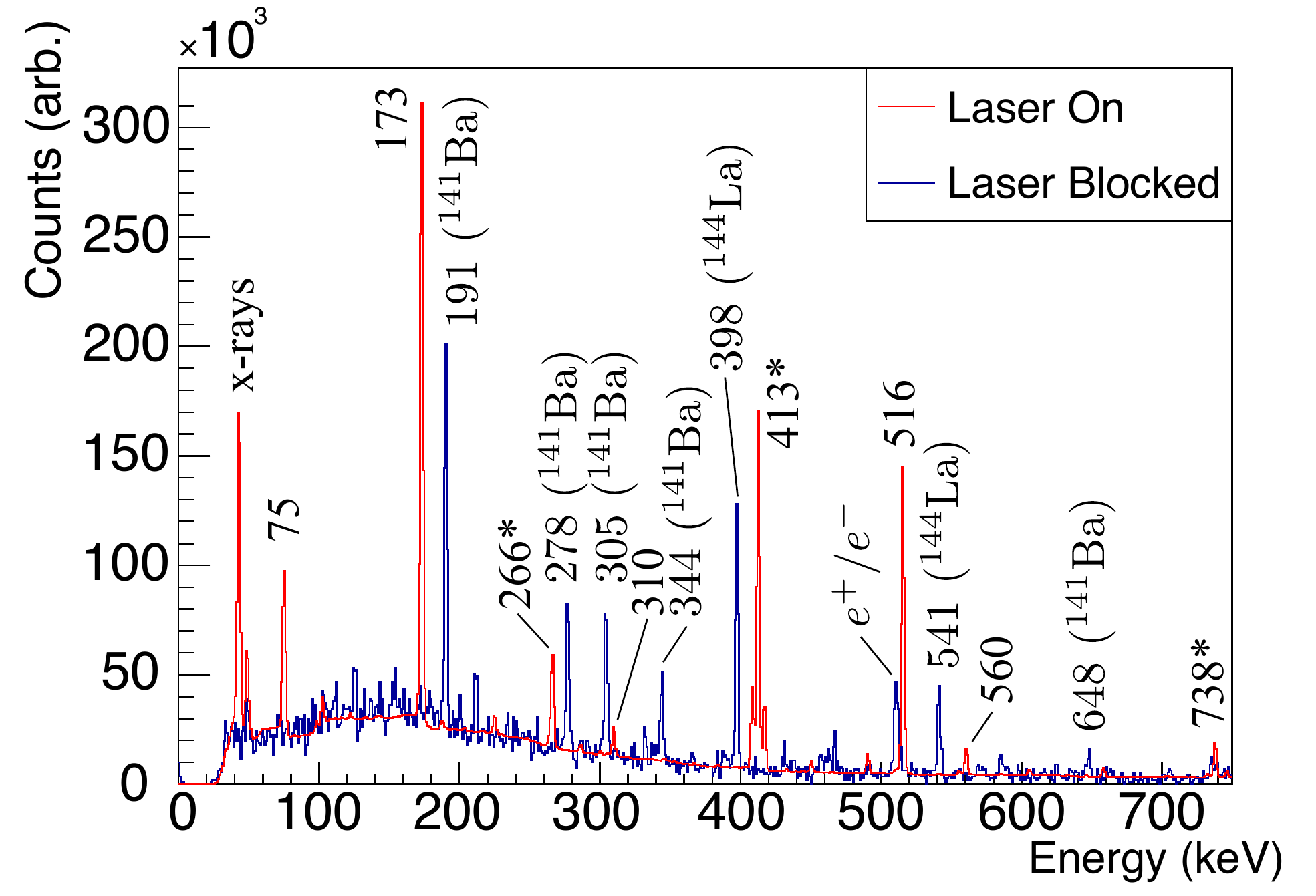}
    \caption{(Color online) Comparison of $\beta$-gated $\gamma$-ray spectra for $^{160}$Eu $\beta$-decay data with laser ionization on and blocked. Peaks with an asterisk (*) denote multiplets. Without laser ionization, molecular contaminants of $^{141}$Ba$^{19}$F and $^{144}$La$^{16}$O were present in the beam. The laser on/blocked spectra have been scaled to have a comparable background level in this figure.}
    \label{fig:laser_comp}
\end{figure}

The beam of $^{160}$Eu$^{g,m}$ was delivered to the GRIFFIN facility and implanted into a mylar tape at the center of the detector array. 
A tape cycle of 10 s tape move and background collection, 380 s of beam delivery, and 114 s of decay (no beam delivery) was used based on the previously known half-life of $t_{1/2}=38 (4)$ s \cite{NDS_A160}. The moving tape collector 
removes the ``contaminated" piece of the tape from the implantation point to behind a lead wall to reduce the background from longer-lived daughter decays and allows to implant onto a clean piece of tape during a new cycle.

The Zero Degree Scintillator (ZDS) and Pentagonal Array for Conversion Electron Spectrometers (PACES) detectors were installed inside the implantation vacuum chamber. The ZDS is a 1 mm thick fast plastic scintillator located just behind the implantation point and is used for $\beta$-tagging and coincident fast-timing measurements. PACES is a conversion electron spectrometer consisting of five cryogenically-cooled lithium-drifted silicon detectors used for conversion electron spectroscopy following the $\beta$-decay.

Surrounding the vacuum chamber is the GRIFFIN array, which consisted of 15 high-purity germanium (HPGe) clover detectors with their front face a radial distance of 11 cm from the implantation point. Each HPGe clover consists of four cryogenically-cooled large-volume germanium crystals used for $\gamma$-ray detection. 
The absolute efficiency of the 15 HPGe detectors in this setup was approximately 31\% at 100 keV and 10\% at 1 MeV for the single crystal mode as determined using standard $^{152}$Eu, $^{60}$Co, $^{56}$Co, and $^{133}$Ba sources. Additionally, there were seven LaBr$_3$(Ce) detectors surrounding the implantation chamber. The LaBr$_3$(Ce) detectors have their signals processed by analog electronics and  Time-to-Amplitude Converters (TACs) to provide excellent timing resolution and are used for fast-timing measurements to determine the lifetimes of excited states in the ps to ns range. More information on the GRIFFIN spectrometer and its components can be found in Refs. \cite{GRIFFIN_NIM,GRIFFIN_DAQ,GRIFFIN_HPGe}.

Approximately 3.9 hours of $^{160}$Eu$^\mathrm{g,m}$ decay data was collected with an intensity of $\sim$3000 pps while using laser ionization. An additional 45 minutes of data was collected with one laser ionization step blocked. By comparing spectra with laser ionization on and blocked, contaminants in the beam can be readily identified. Figure~\ref{fig:laser_comp} compares spectra for both laser settings; with lasers blocked, it can be seen that the beam is contaminated with $^{141}$Ba and $^{144}$La. These species are transmitted through the high-resolution mass separator as molecular $^{141}$Ba$^{19}$F and $^{144}$La$^{16}$O. Based on this observation, we can clearly identify that the $\gamma$-rays observed in the laser-ionized spectra are due to the $^{160}$Eu $\beta$-decays and not contaminants.

The fitting of the number of detected $\beta$-particles in the decay curve using the two measured half-lives of the $\beta$-decaying states (see Sec.~\ref{sec:half_lives}) yields the ratio of ground- and isomeric-states in the $^{160}$Eu beam. 
Our analysis determined that the composition of $^{160}$Eu in the beam contained 70.7(3)\% of high-spin state and 29.3(3)\% of low-spin state.


\section{Results}\label{Sec:Res}

\begin{figure}
    \centering
    \includegraphics[width=1\linewidth]{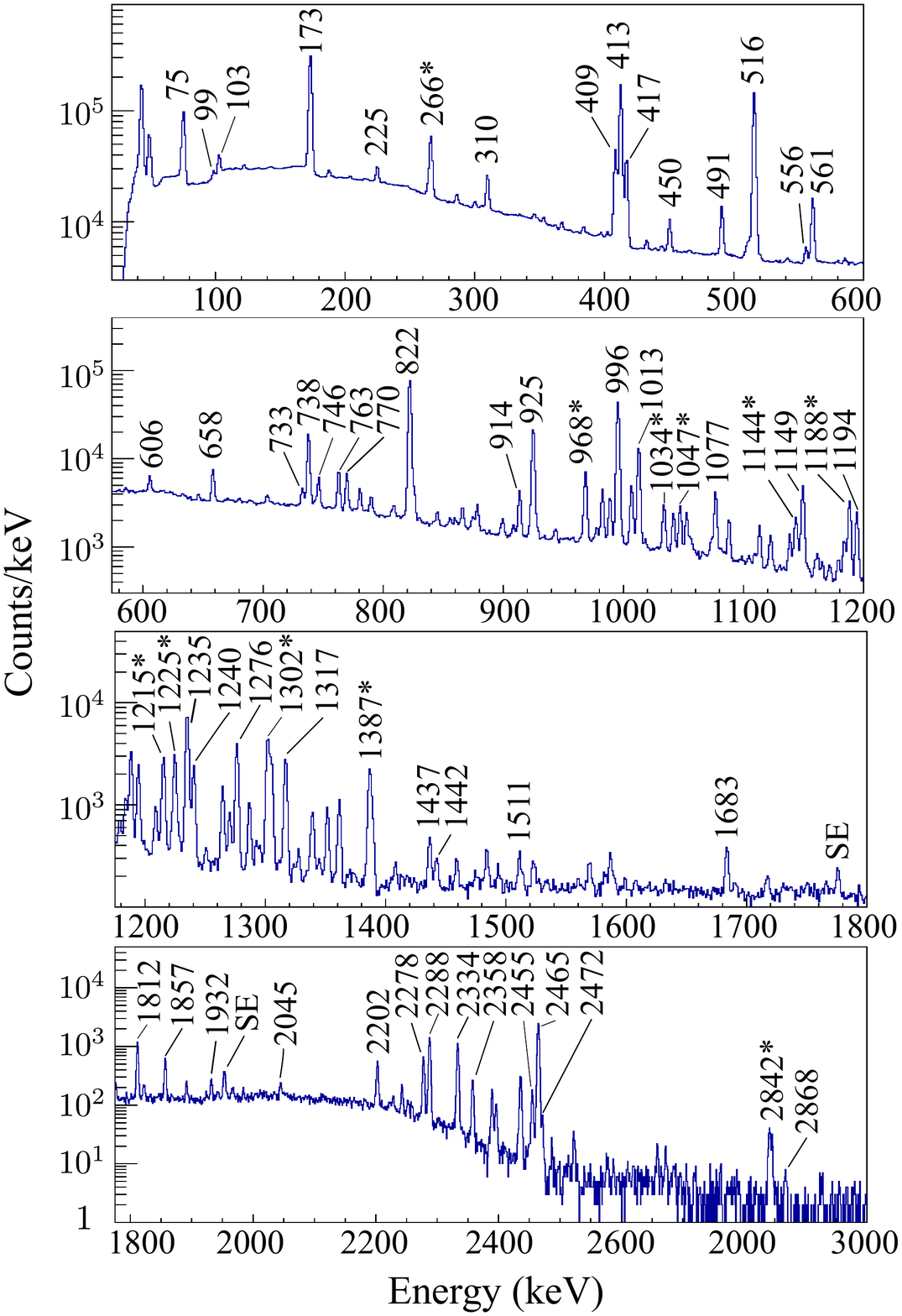}
    \caption{$\beta$-gated $\gamma$-ray spectra resulting from the $\beta$-decay of $^{160}$Eu$^\mathrm{g,m}$. Most, but not all, major peaks have been labeled. Peaks with an asterisk (*) are known multiplets. Single escape peaks have been labeled with SE. 
    }
    \label{fig:gE_z_FER}
\end{figure}

$\beta$-tagged $\gamma$-ray spectra from the $\beta$-decay of $^{160}$Eu are shown in Fig.~\ref{fig:gE_z_FER}. Note that while the Q-value of the ground-state $\beta$-decay is 4448.6(14) keV (+93.0(12) keV for the isomeric-state) \cite{AME2020,Hartley_PRL}, no $\gamma$-rays above 3 MeV were observed. Our sensitivity limit for observation of these high-energy transitions relative to the 173-keV transition for the high-spin decay is $I_{\gamma,\mathrm{rel}}=0.01\%$ at approximately 3 MeV.

Excited states previously identified in the $\beta$-decay study of Ref.~\cite{Hartley_PRL,Hartley_PRC} have been confirmed, and a majority of the decaying transitions were observed. Additionally, two previously unknown excited states (at $E_x=$ 3090 and 3116~keV) and ten new transitions resulting from the high-spin decay were identified in this work. $\gamma$-ray intensities were generally determined through $\gamma$-ray singles spectra, with some weak transitions or doublets determined via $\gamma$-$\gamma$ coincidences.

\begin{figure*}[!htbp!]
\includegraphics[width=1\textwidth]{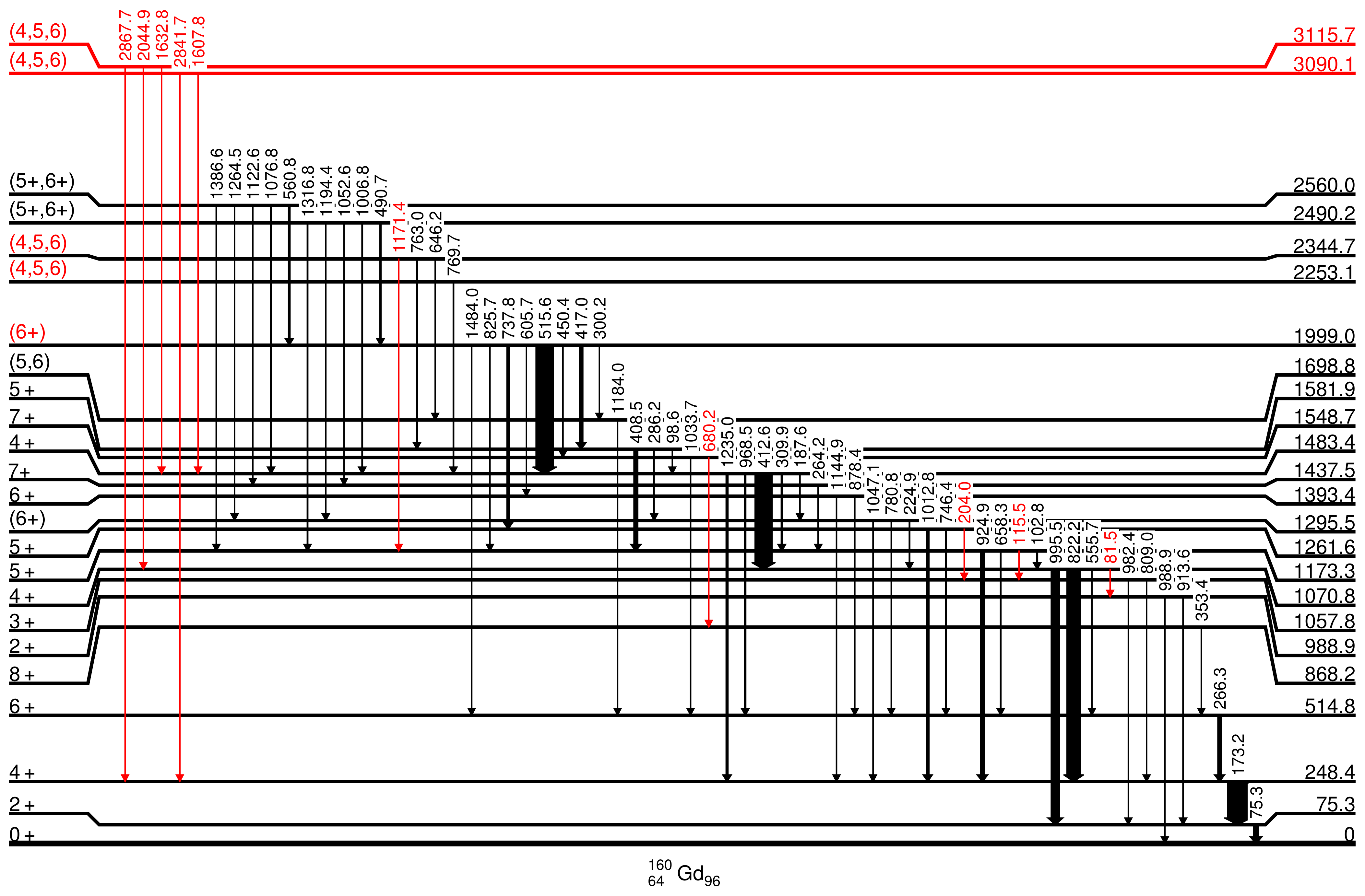}
\caption{\label{fig:level_scheme_high}(Color online) Level scheme of $^{160}$Gd resulting from the $\beta$-decay of the $J$=($5$) high-spin $\beta$-decaying state in $^{160}$Eu. Previously identified levels, transitions, and spin assignments are shown in black, with new or updated ones in red. Arrow widths indicate relative $\gamma$-ray intensities.}
\end{figure*}

Note that because both the high- and low-spin $\beta$-decaying states depopulate through the $2^+$ and $4^+$ members of the ground-state band, it is difficult to determine the respective $\beta$-feeding to them.
While in most cases only upper-bounds on $\beta$-feeding could be determined, no evidence for direct $\beta$-feeding to the ground-state band from the high-spin $\beta$-decaying isomer was observed. This is consistent with the expectation given the high degree of K-forbiddenness of these transitions.

The fraction of 173.2-keV transition associated with the high-spin $\beta$-decay was determined by balancing the transition intensity ($\gamma$+CE) into and out of the $4^+$ state for all transitions coming from the high-spin $\beta$-decay. The remaining intensity of the 173.2-keV transition was then associated with the low-spin $\beta$-decay and used to determine the $\beta$-feeding intensity to the $4^+$ member of the ground-state band. The fraction of 75.3-keV transition from the $2^+$ member of the ground-state band was determined in a similar way: the transition intensities for the high-spin decay were balanced, and the remaining intensity was assigned to the low-spin decay used for determining $\beta$-feeding from the low-spin $\beta$-decay to the $2^+$ state in the ground-state band. It should be noted that these set upper limits on the $\beta$-feeding to these states, as unobserved $\gamma$-ray transitions from both the low- and high-spin $\beta$-decays would decrease the calculated $\beta$-feeding intensities into the $2^+$ and $4^+$ members of the ground-state band.

The 988.9- and 1057.7-keV states are predominantly populated by the low-spin $\beta$-decay, but the weak 81.5-, 115.5-, and 204.0-keV transitions that feed from states populated by the high-spin $\beta$-decay into the 988.9- and 1057.7-keV states make it difficult to determine the $\gamma$-ray intensity of transitions from these states. Similar to above, we have assumed no direct $\beta$-feeding from the high-spin parent to the 988.9-keV state because of its spin of $2^+$. Intensities ($\gamma$-ray+conversion electron) for the populating 81.5-keV transition and depopulating 913.6- and 988.9-keV transitions were balanced to give $I_\beta=0$ to this state, and the remaining $\gamma$-ray intensities of the 913.6- and 988.9-keV transitions were assigned to the low-spin $\beta$-decay. The same procedure was used for the 1057.7-keV state populated by the 115.5- and 204.0-keV transitions and depopulated by the 809.0- and 932.4-keV transitions.

In Refs.~\cite{Hartley_PRL,Hartley_PRC}, the $^{160}$Eu ground-state was assigned spin ($5^-$) and the isomeric-state at 93.0 keV spin ($1^-$). Results from our analysis, discussed in detail in Sec.~\ref{Sec:Disc}, suggest that the high-spin $\beta$-decaying state in $^{160}$Eu is spin $J^\pi=(5)$ and the low-spin $\beta$-decaying is spin ($1$), with no determination as to which is the ground-state. For the purpose of determining quantities such as log($ft$) values in this section, we have used both as being the ground-state, as the difference between log($ft$) values from the ground-state versus the 93.0-keV isomeric-state are generally $\sim$1\% or less.



\begin{longtable}[!pt!]{ccccccc}
    \caption{Decay information for the high-spin $\beta$-decay of $^{160}$Eu. $\gamma$-ray intensities are relative to the 173.2-keV transition. Literature values $I_{\gamma,lit}$ are taken from Ref.~\cite{Hartley_PRC}.}\\
    \hline \hline
    $ E_{level}$ & $J^\pi$ &$I_\beta$ & log($ft$) &$E_\gamma$ & $I_{\gamma,rel}$  & $I_{\gamma,lit}$ \\ 
    (keV) & & (\%) & & (keV) &  & \\\hline
    \endfirsthead
    
    \caption{$(Continued.)$}\\
    \hline \hline
     $ E_{level}$ & $J^\pi$ &$I_\beta$ & log($ft$) &$E_\gamma$ & $I_{\gamma,rel}$  & $I_{\gamma,lit}$ \\ 
    (keV) & & (\%) & & (keV) &  & \\\hline
    
    \endhead
    \hline
    \endfoot
    \hline \hline
    \endlastfoot
       
       75.3(2) & $2^+$        &    -    &    -     & 75.3(2)   & 22(1)   & 23.1(12) \\
       248.4(3) & $4^+$       & -       & -        & 173.2(2)  & 100(4)  & 100(5)\\
       514.8(3) & $6^+$       & -       & -        & 266.3(2)  & 18.3(5) & 17.3(9)\\
       868.2(4) & $8^+$       & -       & -        & 353.4(2)  & 2.70(9) & - \\
       988.9(2) & $2^+$       & -       &   -      & 913.6(2)  & 1.23(7) & -\\
                &             &         &          & 988.9(2)  & 0.46(3) & -\\
      1057.7(3) & $3^+$       & -       &   -      & 809.0(2)  & 0.24(2) & -\\
                &             &         &          & 982.4(2)  & 1.07(7) & -\\
      1070.8(3) & $4^+$       & 1(2)    & $>$7.0   & 81.5(2)   & 0.26(8) & - \\
                &             &         &          & 555.7(2)  & 0.98(5) & 0.97(23) \\
                &             &         &          & 822.2(2)  & 69(2)   & 66.1(33)\\
                &             &         &          & 995.5(2)  & 43(1) & 54.1(27) \\
      1173.3(3) & $5^+$       & 0.4(5)  & $>$7.4   & 102.8(2)  & 5.3(2)  & 6.2(4)\\
                &             &         &          & 115.5(2)  & 0.5(1)  & - \\
                &             &         &          & 658.3(2)  & 3.46(9) & 2.78(19)\\
                &             &         &          & 924.9(2)  & 22.5(5) & 20.7(11)\\
      1261.3(3) & $5^+$       &  1.6(3) & 7.12(1)  & 204.0(2)  & 0.60(5) &  - \\
                &             &         &          & 746.4(2)  & 2.56(8) & 3.08(20) \\
                &             &         &          & 1012.8(2) & 13.4(3) & 13.5(7) \\
      1295.5(3) & ($6^+$)     & 0.0(2)  & $>$7.9   & 224.9(2)  & 3.6(1)  & 4.07(24) \\
                &             &         &          & 780.8(2)  & 2.12(7) & 1.9(4) \\
                &             &         &          & 1047.1(2) & 1.4(3)  & 2.5(8) \\
      1393.4(3) & $6^+$       & 0.2(1)  & 7.85(1)  & 878.4(2)  & 1.2(1)  & 1.1(4) \\
                &             &         &          & 1144.9(2) & 0.7(2)  & 1.5(9) \\
      1437.5(4) & $7^+$       & 0.0(2)  & $>$7.8   & 264.2(2)  & 3.4(2)  & 2.1(3)\\
      1483.4(3) & $4^+$       & 6(2)    & 6.40(1)  & 187.6(2)  & 3.1(1)  & 1.7(6) \\ 
                &             &         &          & 309.9(2)  & 6.4(2)  & 6.4(4) \\
                &             &         &          & 412.6(2)  & 88(2)   & 84.5(43)\\
                &             &         &          & 968.5(2)  & 5.5(3)  & 3.9(5)\\
                &             &         &          & 1235.0(2) & 9.1(2)  & 11.3(6)\\
      1548.5(5) & $7^+$       & 0.1(3)  & $>$7.5   & 680.2(2)  & 0.19(4) & - \\
                &             &         &          & 1033.7(3) & 2.5(6)  & 2.66(18) \\
      1581.9(4) & $5^+$       & 1.7(4)  & 6.88(1)  & 98.6(2)   & 1.23(7) & 1.6(4) \\
                &             &         &          & 286.2(2)  & 1.47(7) & $<$0.25\\
                &             &         &          & 408.5(2)  & 20.2(5) & 17.9(9)\\
      1698.8(4) & (5,6)       & 0.06(5) & 8.271)  & 1184.0(2) & 1.21(5) & 1.5(4) \\
      1999.0(4) & $(6^+)$     & 63(2)   & 5.04(1)  & 300.2(2)  & 0.70(5) & 0.49(20)\\
                &             &         &          & 417.0(2)  & 18.3(5) & 15.1(8) \\
                &             &         &          & 450.4(2)  & 2.59(8) & 2.0(4)\\
                &             &         &          & 515.6(2)  & 89(2)   & 87.9(44) \\
                &             &         &          & 605.7(2)  & 1.50(8) & 2.2(5)\\
                &             &         &          & 737.8(2)  &13.2(3) & 14.4(8) \\
                &             &         &          & 825.7(2)  & 1.98(7) & 1.96(18) \\
                &             &         &          & 1484.0(2) & 0.25(2) & 0.56(26) \\
      2253.1(4) &  ($4,5,6$)  & 2.31(9) & 6.29(1)  & 769.7(2)  & 4.3(1)  & 4.20(25)\\
      2344.7(4) &  ($4,5,6$)  & 2.8(1)  & 6.13(1)  & 646.2(2)  & 0.33(4) & 0.25(13) \\
                &             &         &          & 763.0(2)  & 4.3(1)  & 4.14(25) \\
                &             &         &          & 1171.4(3) & 0.46(7) & - \\
      2490.2(4) & ($5^+,6^+$) & 9.2(4)  & 5.49(1)  & 490.7(2)  & 5.3(1)  & 5.1(3) \\
                &             &         &          & 1006.8(2) & 3.8(1)  & 3.55(22)\\
                &             &         &          & 1052.6(2) & 1.1(4)  & 0.85(12)\\
                &             &         &          & 1194.4(2) & 2.67(8) & 0.25(9)\\
                &             &         &          & 1316.8(2) & 3.9(1) & 3.20(20)\\
      2560.0(4) & ($5^+,6^+$) & 11.3(4) & 5.34(1)  & 560.8(2)  & 8.6(2)  & 7.4(4) \\
                &             &         &          & 1076.8(2) & 3.8(1)  & 4.2(3) \\
                &             &         &          & 1122.6(2) & 3.23(9) & 0.70(8)\\
                &             &         &          & 1264.5(2) & 1.51(5) & 1.18(16) \\
                &             &         &          & 1386.6(2) & 3.43(9) & 3.1(4)\\
      3090.1(4) & ($4,5,6$)   & 0.16(2) & 6.64(1)  & 1607.8(4) & 0.02(2) & -\\
                &             &         &          & 2841.7(2) & 0.27(2) & - \\
      3115.7(3) & ($4,5,6$)   & 0.41(3) & 6.20(1)  & 1632.8(4) & 0.43(3) & - \\
                &             &         &          & 2044.9(2) & 0.29(2) & - \\
                &             &         &          & 2867.7(5) & 0.03(1) & \;- 

    \label{tab:HS_gammas}
\end{longtable}





\begin{figure}[!h!]
    \centering
    \includegraphics[width=0.85\columnwidth]{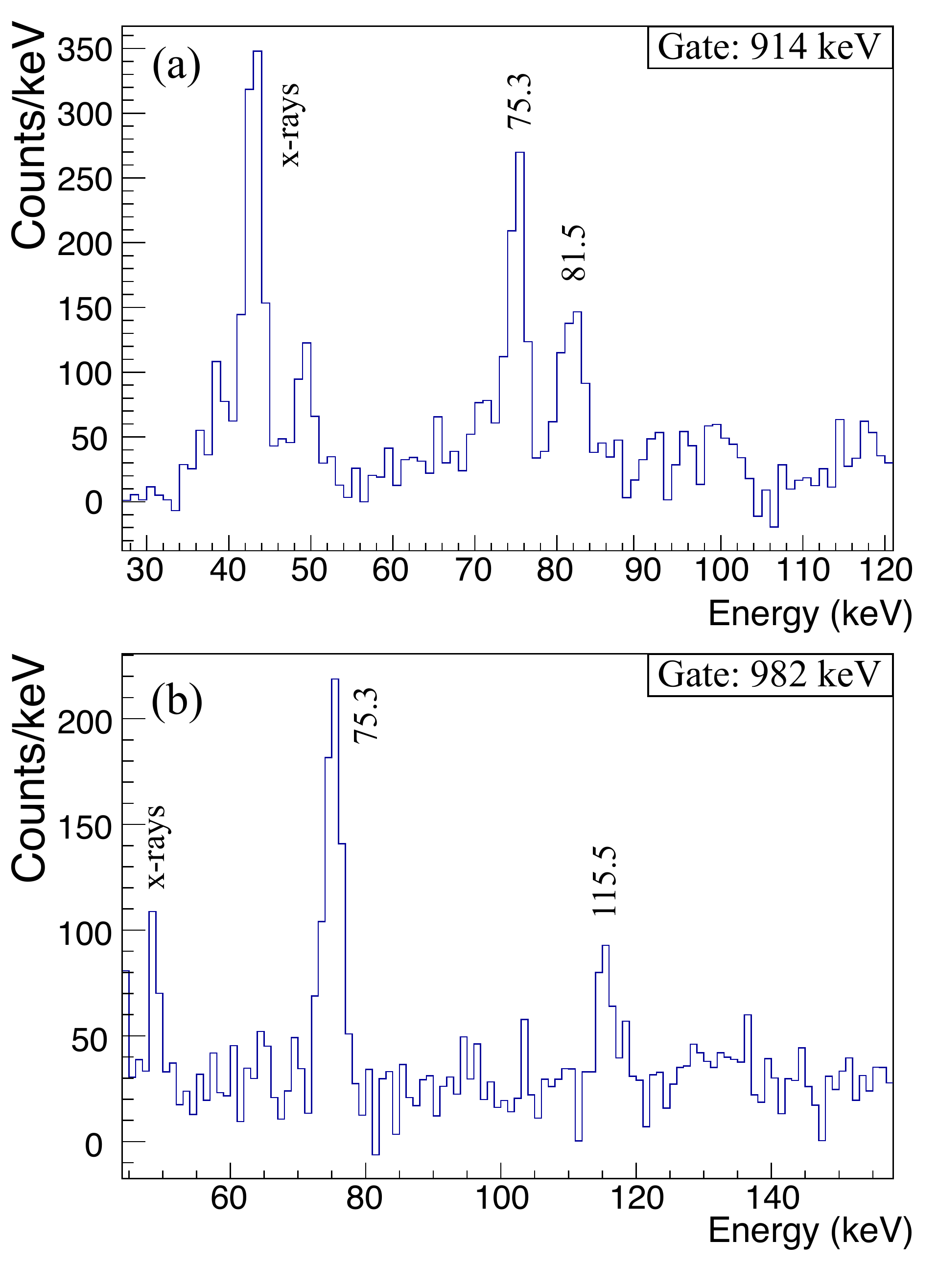}
    \caption{$\beta$-$\gamma$-gated spectra showing the (a) 81.5 keV and (b) 115.5 keV $\gamma$-rays. These transitions connect the first $K^\pi$=$4^+$ band to the $\gamma$-band and are only observed in coincidence spectra when gating on a depopulating transition.}
    \label{fig:82_116_gammas}
\end{figure}

\subsection{Excited States Populated by the $J$=($5$) $\beta$-decaying State}

The $\gamma$-rays associated with the de-excitation of the high-spin states in $^{160}$Gd were placed in the level scheme as shown in Fig.~\ref{fig:level_scheme_high}. Decay information for levels and transitions from the high-spin $\beta$-decay are shown in Table~\ref{tab:HS_gammas}. The $\gamma$-ray intensities were corrected for summing effects using the method outlined in Ref.~\cite{GRIFFIN_NIM}. 
The $\beta$-feeding intensities were determined by balancing transition intensities populating and depopulating each state and were normalized to 100\% for both the high-spin and low-spin $\beta$-decays. For states where the $\beta$-feeding intensity was calculated to be less than 0, the intensity was set to 0 prior to normalization of the $\beta$-feeding intensities.

Conversion electron corrections were applied using BrIcc V2.3 \cite{BRICC} for $\gamma$-rays below 600 keV with $\alpha>0.01$ when determining $\beta$-feeding intensities. For transitions between states without firm spin assignments, the transition was assumed to be the lowest multipole -- generally $M1$ -- for the purpose of conversion electron corrections. Log($ft$) values were calculated based on the $\beta$-feeding intensities and the determined half-life of the $\beta$-decaying state (see Sec.~\ref{sec:half_lives}). For states where the $\beta$-feeding uncertainty is equal to or larger than the intensity, lower limits for the log($ft$) values have been provided.

In the following, we highlight discrepancies and new results compared with those published in Ref.~\cite{Hartley_PRC}.

\subsubsection{868.2 keV-state}
We observe the $8^+$ member of the ground-state band at 868.2 keV with its associated 353.4-keV depopulating transition. While this level and transition have been known previously, this is the first observation of them resulting from the $\beta$-decay of $^{160}$Eu.

\subsubsection{1070.8-keV state}
Of interest is the first observation of a low-intensity, low-energy $\gamma$-ray from this state into the $\gamma$-band. This 81.5 keV transition is only observed when gating from below due to its weak intensity and location in an area of high Compton background. The spectrum resulting from a coincidence condition placed on the 913.6-keV $\gamma$-ray transition can be seen in Fig.~\ref{fig:82_116_gammas}. From this spectrum, a relative intensity of $I_\gamma=0.26(8)$ was determined. Although this transition is highly converted ($\alpha=7.314$ \cite{BRICC}), the low $\gamma$-ray intensity and low conversion electron detection efficiency resulted in no observed conversion electron signal for this transition in PACES.

\subsubsection{1173.3-keV state}
As with the 1070.8-keV state, we observed another weak transition at 115.5 keV which connects the $K^\pi$=$4^+$ band and the $\gamma$-band. This transition can be observed in the $\beta$-$\gamma$-gated spectrum in Fig.~\ref{fig:82_116_gammas}. As with the 81.5-keV transition, no associated conversion electron was observed, and a relative intensity of $I_\gamma=0.5(1)$ was determined.

\begin{figure}[htbp]
\centering
    \includegraphics[width=0.85\columnwidth]{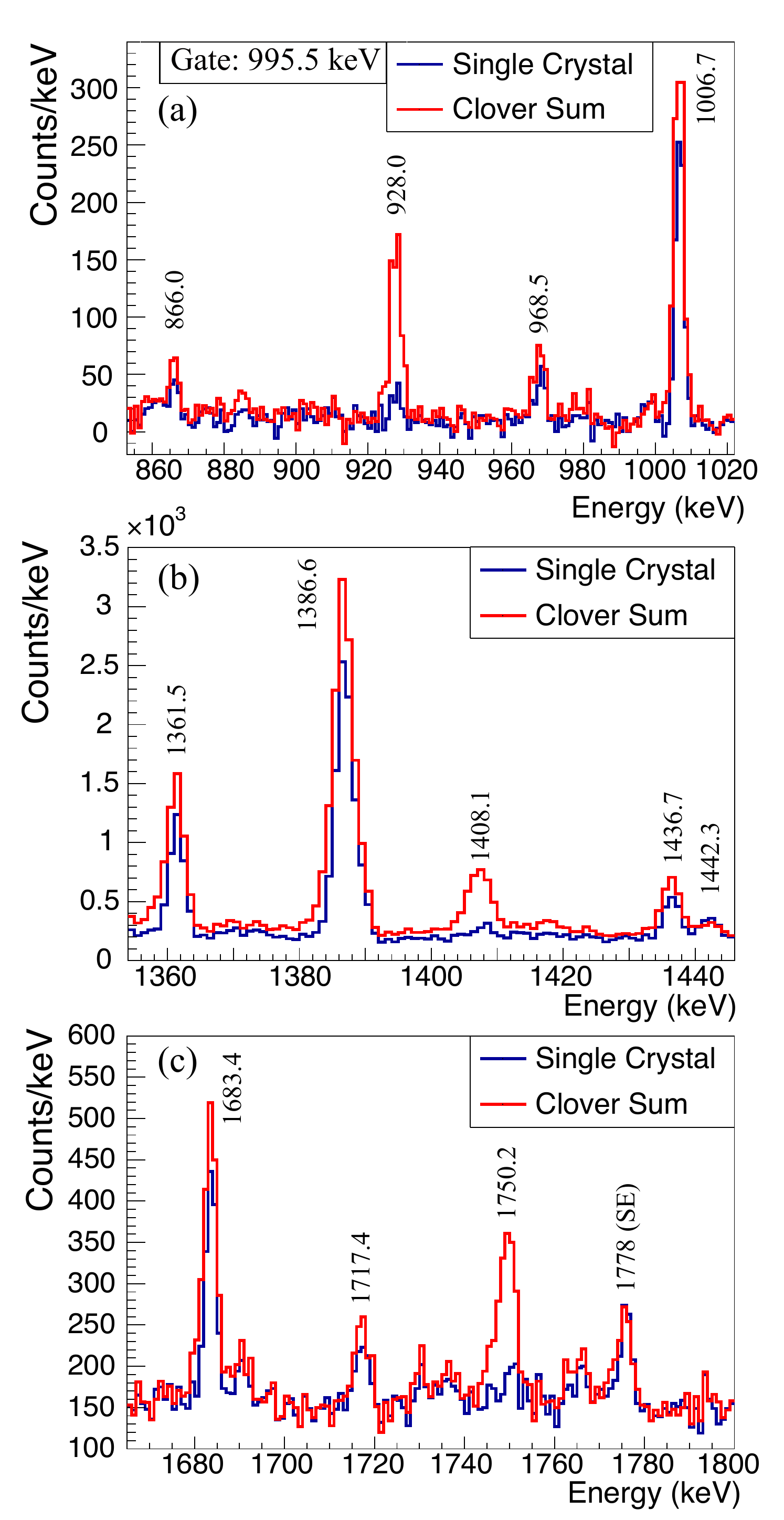}
    \caption{(Color online) Comparison of single crystal and clover sum modes for the (a) 928-keV (gated on the 995.5-keV transition), (b) 1408-keV, and (c) 1750-keV peaks. The significant increase in photopeak area in clover sum mode is characteristic of summed peaks. The 1778-keV peak in (c) is a single escape peak (SE).}
    \label{fig:summed_peaks}
\end{figure}

\begin{figure*}[!htbp!]
\includegraphics[width=1\textwidth]{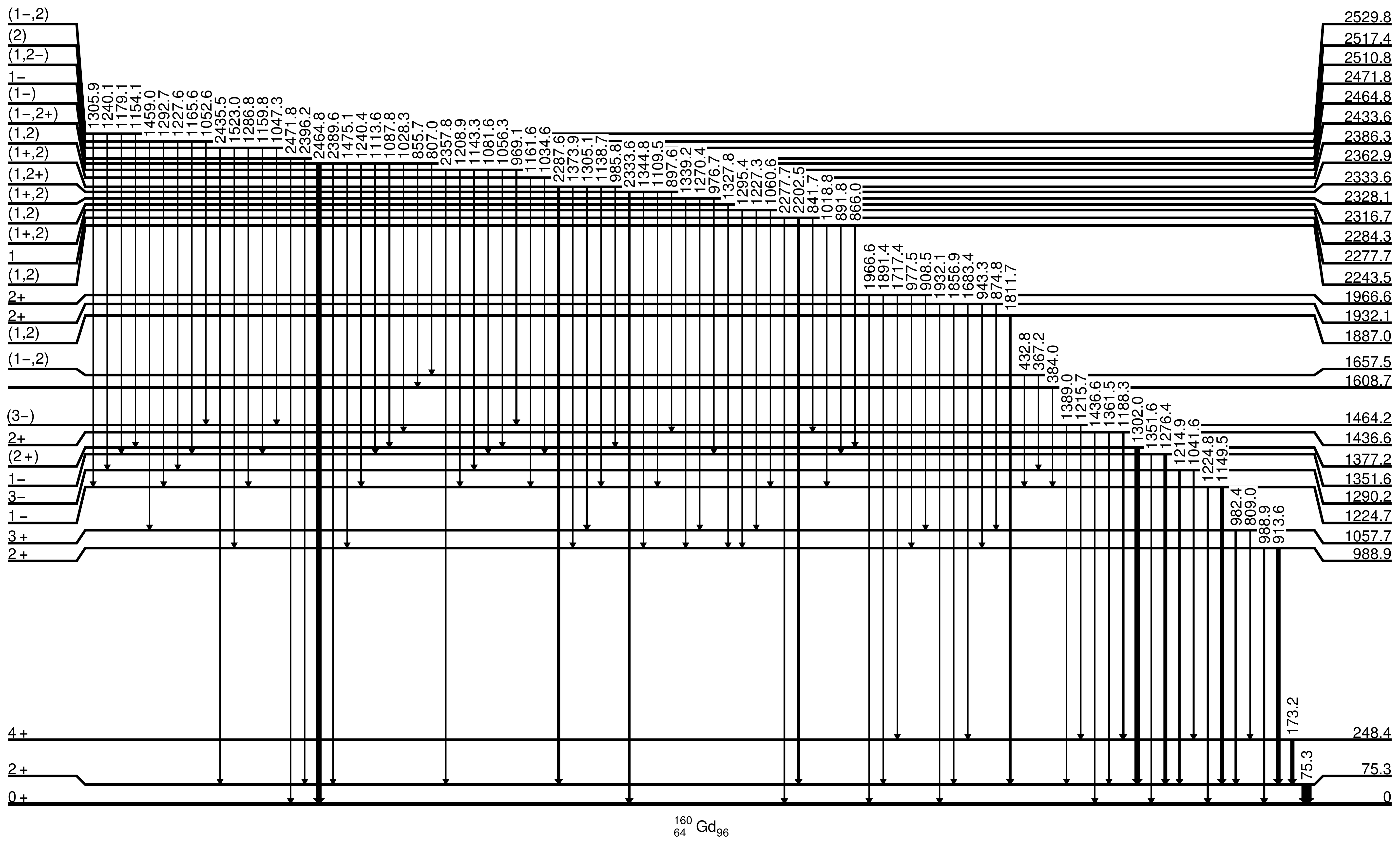}
\caption{\label{fig:level_scheme_low}Level scheme of $^{160}$Gd resulting from the $\beta$-decay of the $J$=($1$) low-spin $\beta$-decaying state in $^{160}$Eu. No new excited states or levels were observed, but some transitions are disputed and not included. Transition widths indicate $\gamma$-ray intensities.}
\end{figure*}

\subsubsection{1261.3-keV state}
We observed a 204.0-keV transition depopulating this state in addition to the 746.4- and 1012.8-keV $\gamma$-rays.

\subsubsection{1295.5-keV state}
Our analysis cannot place the tentative 123-keV transition from this state seen in Ref.~\cite{Hartley_PRC}. While we observed a peak at this energy, its abnormal peak shape and lack of coincidences with expected $\gamma$-rays in the level scheme do not justify a placement.

\subsubsection{1483.4-keV state}
We confirm six of the seven $\gamma$-rays depopulating this level observed by Hartley \textit{et al.}~\cite{Hartley_PRC}. In our analysis, the 1408.1-keV peak placed as a depopulating $\gamma$-ray from this level was determined to be due to summing effects from the intense 412.6 -- 995.5-keV cascade. When using the clover-sum mode, where the energies of all four crystals within a single HPGe clover for an event are summed, peaks due to summing show a significant increase in area compared to neighboring real photopeaks (Fig.~\ref{fig:summed_peaks}). After correcting for summing effects, 
the 1408.1-keV $\gamma$-ray was determined to have a negative intensity -- another indication that it is due to summing and not a real transition. Ref.~\cite{Hartley_PRC} used the clover-sum mode in their analysis, so we speculate that their observation of this $\gamma$-ray is due to summing of the intense transitions within the HPGe clovers.


\subsubsection{1548.5-keV state}
Hartley \textit{et al.}~\cite{Hartley_PRC} place a 286.9-keV $\gamma$-ray as depopulating this state. Our analysis cannot confirm the placement of this $\gamma$-ray as a depopulating transition from the 1548.5-keV state. However, we do observe a 680.2-keV $\gamma$-ray from this state that populates the known $8^+$ member of the ground-state band at an energy of 868.2 keV. 

\subsubsection{1698.8-keV state}

We did not observe the tentative 215-keV $\gamma$-ray from this state feeding the 1483.4-keV state as published in Ref.~\cite{Hartley_PRC}.

\subsubsection{1999.0-keV state}
We observe eight of the ten previously-observed decaying $\gamma$-rays from this state. The 928.0- and 1750.2-keV $\gamma$-rays observed by Ref.~\cite{Hartley_PRC} were determined to be due to clover summing effects rather than true photopeaks. The increase in peak area associated with summing using the clover sum mode can be seen from Fig.~\ref{fig:summed_peaks}. As with the 1408.1-keV $\gamma$-ray, the intensities for both the 928.0- and 1750.2-keV $\gamma$-rays were determined to be less than 0\% after summing corrections were applied.

The spin of this state was assigned ($5^-$) in Refs.~\cite{Hartley_PRL,Hartley_PRC}. Our analysis, however, assigns a spin of ($6^+$). A detailed justification of this spin assignment can be found in Sec.~\ref{Sec:Disc}.

\subsubsection{2344.7-keV state}

We confirm the tentatively-placed 646-keV $\gamma$-ray \cite{Hartley_PRC} depopulating this level, as well as a 1171.4-keV $\gamma$-ray. Based on the decay pattern, the spin is likely $(4,5,6)$.

\subsubsection{3090.1-keV state}
Our analysis reveals a new level at 3090.1-keV with two depopulating $\gamma$-rays of 1607.8- and 2841.7-keV. The state is likely of spin 4, 5, or 6 given its transitions to two $4^+$ states and the spin($5$) assignment of the high-spin parent in $^{160}$Eu.

\subsubsection{3115.7-keV state}
Another new state at 3115.7 keV was observed. Three new $\gamma$-rays of energies 1632.8, 2044.9, and 2867.7 keV are depopulating this level. As with the 3090.1-keV state, the only $\gamma$-rays that decay from this level are seen to populate $4^+$ states, so it is likely of spin 4, 5, or 6.



\subsection{Excited States Populated by the $J$=($1$) $\beta$-decaying State}

Our analysis confirms all levels published by Hartley \textit{et al.}~\cite{Hartley_PRC} and the majority of the associated $\gamma$-rays; discrepancies in the presence of transitions are listed below. While our analysis largely agrees with the placement of transitions, some $\gamma$-ray intensities vary between our analysis and that of Ref.~\cite{Hartley_PRC}. Branching ratios of the depopulating $\gamma$-rays are generally consistent with previous literature values. 

The respective decay information is shown in Table~\ref{tab:LS_gammas} and the level scheme in Fig.~\ref{fig:level_scheme_low}. As with the high-spin $\beta$-decay, $\beta$-intensities are normalized to the low-spin $\beta$-decay, and excited-states with a $\beta$-feeding intensity less than 0 were set to 0 prior to the $\beta$-feeding normalization. Because we were unable to determine the $\beta$-feeding direct to the ground-state, $\beta$-feeding intensities should be treated as upper limits, and no log($ft$) values are provided.


\begin{longtable}[!t!]{cccccc}
    \caption{Decay information for the low-spin $\beta$-decay of $^{160}$Eu. $\beta$-feeding intensities are normalized to the low-spin states only. $\gamma$-ray intensities are relative to the 2464.-keV transition. Literature values are taken from Ref.~\cite{Hartley_PRC}.}\\
    \hline \hline
   $ E_{level}$ & $J^\pi$ &$I_\beta$  &$E_\gamma$ & $I_{\gamma,rel}$  & $I_{\gamma,lit}$ \\ 
    (keV) & & (\%) & (keV) & &  \\\hline
     \endfirsthead
    \caption{$(Continued.)$}\\
    \hline \hline
    $ E_{level}$ & $J^\pi$ &$I_\beta$ &$E_\gamma$ & $I_{\gamma,rel}$  & $I_{\gamma,lit}$ \\ 
    (keV) & & (\%) & (keV) & & \\\hline
    
    \endhead
    \hline
    \endfoot
    \hline \hline
    
    \endlastfoot
       0         & $0^+$       & -       &  -        &  -      & -\\
       75.3(2)   & $2^+$       & 50(9)   & 75.3(2)   & 212(14) & -\\
       248.4(3)  & $4^+$       & 3(3)    & 173.2(2)  & 128(36) & -\\
       988.9(2)  & $2^+$       & 2.2(3)  & 913.6(2)  & 66(1)   & 26.07(48)\\
                 &             &         & 988.9(2)  & 24(1)   & 26.81(52)\\
       1057.7(3) & $3^+$       & 0.0(4)  & 809.0(2)  & 6.2(8)  & 8.04(37)\\
                 &             &         & 982.4(2)  & 28(1)   & 37.96(56)\\
       1224.7(2) & $1^-$       & 0.0(4)  & 1149.5(2) & 66(2)   & 77.22(56)\\
                 &             &         & 1224.7(2) & 27(4)   & 49.44(63)\\
       1290.2(3) & $3^-$       & 0.0(3)  & 1041.6(2) & 23.1(8) & 19.04(44)\\
                 &             &         & 1214.9(2) & 32(2)   & 32.80(150)\\
       1351.6(2) & $1^-$       & 0.6(3)  & 1276.4(2) & 65(2)   & 35.20(160) \\
                 &             &         & 1351.6(2) & 13.4(6) & 6.30(30)\\
       1377.2(3) & ($2^+$)     & 2.0(4)  & 1302.0(2) & 99(7)   & 83.30(110)\\
       1436.6(2) & $2^+$       & 2.2(3)  & 1188.3(2) & 43(1)   & 45.04(59)\\
                 &             &         & 1361.5(2) & 15.3(6) & 17.74(41)\\
                 &             &         & 1436.6(2) & 6.6(4)  & 2.85(15)\\
       1464.2(3) & ($3^-$)     & 0.0(2)  & 1215.7(2) & 16(1)   & 26.10(120)\\
                 &             &         & 1389.0(2) & 10.6(5) & 16.74(37)\\
       1608.7(3) &             & 0.30(5)  & 384.0(2) & 9.8(7)  & 9.19(37)\\
       1657.5(3) & ($1^-,2$)   & 0.55(7) & 367.2(2)  & 7.5(6)  & 8.67(37)\\
                 &             &         & 432.8(2)  & 7.7(6)  & 9.12(59)\\
       1887.0(3) & ($1,2$)     & 1.3(1)  & 1811.7(2) & 26.4(8) & 24.26(44)\\
       1932.1(2) & $2^+$       & 1.6(2)  & 874.8(2)  & 7(1)    & 5.85(37)\\
                 &             &         & 943.4(2)  & 3.6(5)  & 2.93(26)\\
                 &             &         & 1683.4(2) & 5.8(3)  & 7.67(30)\\
                 &             &         & 1856.9(2) & 12.0(4) & 11.78(33)\\
                 &             &         & 1932.1(2) & 4.0(3)  & 4.26(26)\\
       1966.6(3) & $2^+$       & 1.0(2)  & 908.5(2)  & 2.7(4)  & 3.85(37)\\
                 &             &         & 977.5(2)  & 9(3)    & 2.15(19)\\
                 &             &         & 1717.4(2) & 1.4(2)  & 2.00(19)\\
                 &             &         & 1891.4(2) & 3.5(3)  & 2.19(19) \\
                 &             &         & 1966.6(3) & 2.1(3)  & 6.26(30) \\
       2243.5(3) & ($1,2$)     & 1.4(2)  & 866.0(2)  & 21(1)   & 0.41(7)\\
                 &             &         & 891.8(2)  & 5(1)    & 1.04(11)\\
                 &             &         & 1018.8(2) & 1.8(4)  & 2.78(26)\\
       2277.7(2) & $1$         & 3.6(4)  & 841.7(2)  & 17(4)   & 1.41(15) \\
                 &             &         & 2202.5(2) & 34(1)   & 15.26(41) \\
                 &             &         & 2277.7(2) & 19.7(7) & 20.70(44)\\
       2284.3(3) & ($1^+,2$)   & 1.1(2)  & 1060.6(2) & 5(2)     & 1.04(11)\\
                 &             &         & 1227.3(2) & 8(3)    & 0.81(11)\\
                 &             &         & 1295.4(2) & 7.3(6)  & 1.70(19)\\
       2316.7(3) & ($1,2$)     & 0.07(1) & 1327.8(2) & 1.4(3)  & 1.56(19)\\
       2328.1(3) & ($1^+,2$)   & 1.1(1)  & 976.7(3)  & 4(1)    & 1.15(11) \\
                 &             &         & 1270.4(2) & 7.6(4)  & 9.19(33) \\
                 &             &         & 1339.2(2) & 11.4(6) & 13.74(37)\\
       2333.6(2) & ($1,2^+$)   &  2.2(2) & 897.6(2)  & 1.9(8)  & 0.74(7) \\ 
                 &             &         & 1109.5(2) & 3.1(5)  & 3.67(37) \\
                 &             &         & 1344.8(2) & 1.4(4)  & 1.11(15) \\
                 &             &         & 2333.6(2) & 38(1)   & 0.15(4)\\
       2362.9(3) & ($1^+,2$)   & 4.5(5)  & 985.8(3)  & 3.1(8)  & 0.81(11)\\
                 &             &         & 1138.7(2) & 12(1)   & 11.33(37)\\
                 &             &         & 1305.1(2) & 22(6)   & 5.67(37)\\
                 &             &         & 1373.9(4) & 6(2)    & 0.67(7)\\
                 &             &         & 2287.6(2) & 47(1)   & 44.78(63)\\
       2386.3(3) & ($1,2$)     & 0.35(5) & 1034.6(2) & 1.9(8)  & 2.63(19) \\
                 &             &         & 1161.6(2) & 5.1(5)  & 5.26(37)\\
       2433.6(3) & ($1^-,2^+$) & 4.1(4)  & 969.1(2)  & 26(3)   & 3.33(22)\\
                 &             &         & 1056.3(2) & 6(1)    & 4.60(45)\\
                 &             &         & 1081.6(2) & 8(2)    & 1.80(19) \\
                 &             &         & 1143.3(2) & 21(1)   & 20.81(44)\\
                 &             &         & 1208.9(2) & 11.9(7) & 8.07(33)\\
                 &             &         & 2357.8(2) & 9.5(4)  & 11.78(33)\\
       2464.8(2) & ($1^-$)     & 9.3(8)  & 807.0(2)  & 4.8(8)  & 1.85(15) \\
                 &             &         & 855.7(2)  & 4.1(5)  & 4.48(30)\\
                 &             &         & 1028.3(3) & 1.3(5)  & 0.96(7)\\
                 &             &         & 1087.8(2) & 25(1)   & 18.89(44) \\
                 &             &         & 1113.6(2) & 27(1)   & 14.59(44)\\
                 &             &         & 1240.4(2) & 20(5)   & 22.60(160)\\
                 &             &         & 1475.1(4) & 1.4(3)  & 1.22(7) \\
                 &             &         & 2389.6(2) & 5.0(6)  & 6.63(26)\\
                 &             &         & 2464.8(2) & 100(3)  & 100(11) \\
       2471.8(2) & $1^-$       & 0.25(4) & 2396.2(2) & 3.4(7)  & 1.41(11)\\
                 &             &         & 2471.8(2) & 1.7(3)  & 4.85(22)\\
       2510.8(4) & ($1,2^-$)   & 2.8(3)  & 1047.3(2) & 18(3)   & 7.70(44)\\
                 &             &         & 1159.8(2) & 11(3)   & 2.63(19) \\
                 &             &         & 1286.8(2) & 14.1(6) & 15.33(37)\\
                 &             &         & 1523.0(2) & 4.1(3)  & 1.44(15) \\
                 &             &         & 2435.5(4) & 9.9(8)  & 10.37(33)\\
       2517.4(3) & ($2$)       & 1.9(3)  & 1052.7(2) & 11(5)   & 11.11(41)\\
                 &             &         & 1165.6(2) & 3.1(4)  & 5.11(52) \\
                 &             &         & 1227.6(2) & 10(2)   & 4.56(48)\\
                 &             &         & 1292.7(2) & 4.9(7)  & 9.33(74)\\
                 &             &         & 1460.1(2) & 9(1)    & 4.22(37)\\
       2529.8(3) & ($1^-,2$)   & 1.7(3)  & 1154.1(2) & 2.0(6)  & 0.85(11) \\
                 &             &         & 1179.1(2) & 5.4(5)  & 2.81(19) \\
                 &             &         & 1240.1(2) & 16(4)   & 12.59(74)\\
                 &             &         & 1305.9(2) & 10(3)   & 1.89(15)
    \label{tab:LS_gammas}
\end{longtable}

\subsubsection{1224.7-keV state}
We cannot confirm the placement of the 235.8-keV $\gamma$-ray depopulating this level.

\subsubsection{1377.2-keV state}
Our analysis observes a single $\gamma$-ray at 1302.0-keV depopulating this level. We are unable to confirm the 319.3- and 1128.3-keV transitions depopulating this state.

\subsubsection{1887.0-keV state}
We confirm the 1887.0 level and the associated 1811.7-keV transition. We do not observe the 898.2-keV $\gamma$-ray associated with this level.

\subsubsection{2277.7-keV state}
We do not see the depopulating 1288.9-keV transition from this state, but we can confirm the other three published transitions (841.7, 2202.5, and 2277.7 keV).

\subsubsection{2333.6-keV state}
Four of the five transitions from this state were confirmed; only the 982.5-keV transition cannot be confidently placed here. 

Ref.~\cite{Hartley_PRC} lists the intensity of the 2333.6-keV transition at 0.15\% relative to the 2464.8-keV transition. We observe it at a relative intensity of 38\%. The inset of Fig.~2 in Ref.~\cite{Hartley_PRC} shows the 2333.6-keV $\gamma$-ray at a significant fraction of the intensity of the 2464.8-keV $\gamma$-ray, leading us to conclude that the relative intensity of the 2333.6-keV $\gamma$-ray reported in Table~III of Ref.~\cite{Hartley_PRC} to be an error.

\subsubsection{2362.9-keV state}
Our analysis was unable to place the 898.4-keV transition published in Hartley \textit{et al.}~\cite{Hartley_PRC} originating from this state. All six other transitions depopulating this state can be confirmed.

\subsubsection{2433.6-keV state}
We were unable to place the 2432.9-keV ground-state transition seen by Hartley \textit{et al.} \cite{Hartley_PRC}.

\subsection{Half-lives of $^{160}$Eu$^{g,m}$\label{sec:half_lives}}

The half-lives of the two known $\beta$-decaying states in $^{160}$Eu were determined by examining the time distribution of $\beta$-gated $\gamma$-rays associated with states in $^{160}$Gd populated by the two different spin states in the parent nucleus. For each state, the decay portion of the beam cycle was fit with an exponential and the half-life extracted from the fit. The time distribution histograms are shown in Fig.~\ref{fig:decays_half_lives}.

\subsubsection{High-spin isomer}

The high-spin $J$=($5$) $\beta$-decaying isomer in the parent $^{160}$Eu predominately feeds the 1999.0-keV state in $^{160}$Gd. The time distribution of $\gamma$-rays associated with the depopulation were summed together; 21 $\gamma$-rays (309.9, 412.6, 450.4, 490.7, 515.6, 555.7, 560.8, 737.8, 746.4, 780.8, 822.2, 924.9, 995.5, 1006.8, 1012.8, 1076.8, 1122.6, 1194.4, 1235.0, 1264.5, and 1316.8 keV) were used to fit the decay curve and extract a half-life (top histogram in Fig.~\ref{fig:decays_half_lives}). The resulting half-life of the high-spin state was determined to be $t_{1/2}=42.5(7)$ s. This is in very good agreement with the previously measured half-life of $t_{1/2}=42.6(5)$ s using the 412.6, 515.6, 822.2, and 995.5 keV $\gamma$-rays \cite{Hartley_PRL}.

\subsubsection{Low-spin isomer}

\begin{figure}
    \centering
    \includegraphics[width=1\linewidth]{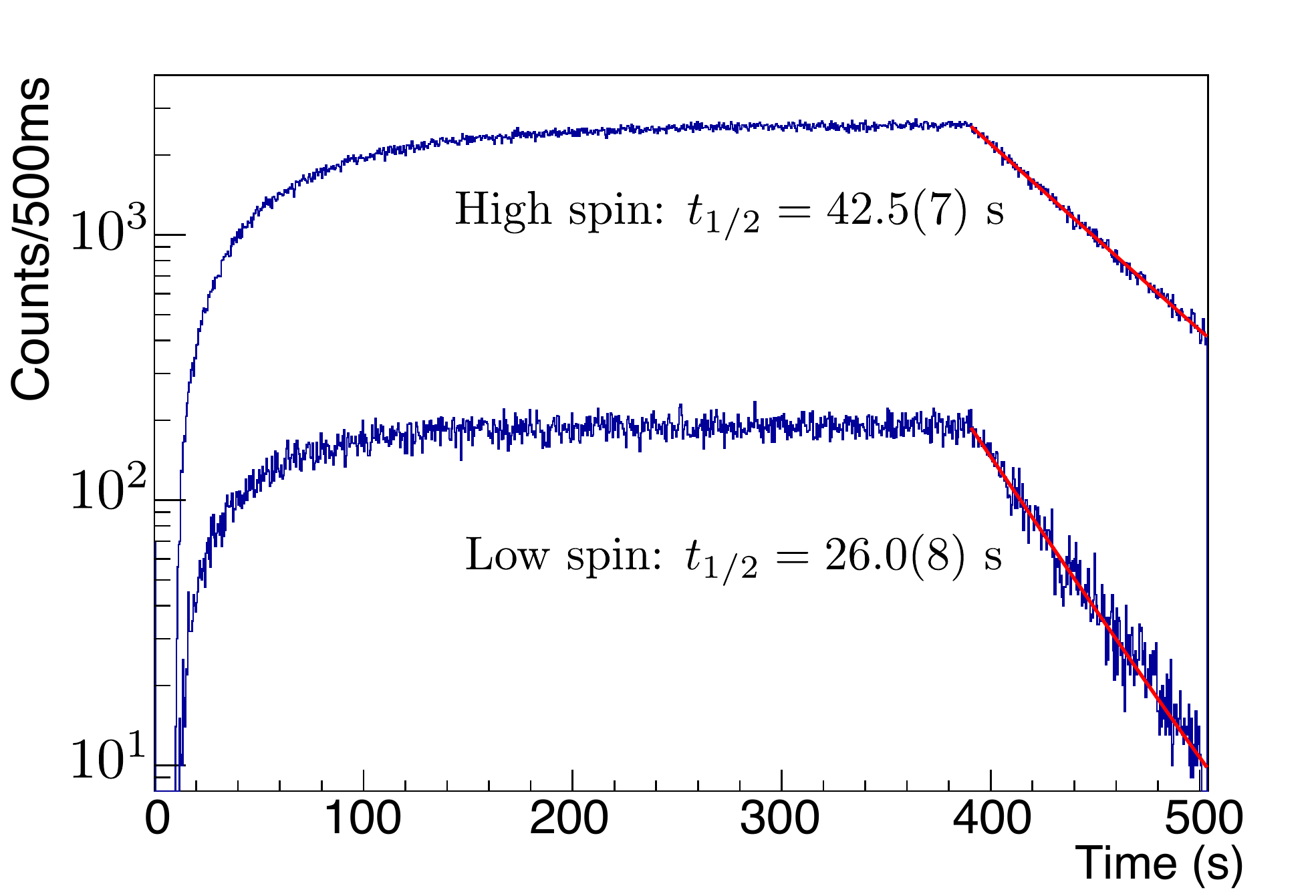}
    \caption{Time distributions of $\gamma$-rays associated with the $\beta$-decay of $^{160}$Eu. The top spectrum shows $\gamma$-rays resulting from the high-spin $\beta$-decay and gives a half-life of $t_{1/2}=42.5(7)$ s, while the bottom spectrum shows the $\gamma$-rays associated with the low-spin decay and has a half-life of $t_{1/2}=26.0(8)$ s.}
    \label{fig:decays_half_lives}
\end{figure}

The decay of the low-spin $\beta$-decaying $J$=($1$) state in $^{160}$Eu is highly fragmented and feeds states of spins $J^\pi$= 1, 2, and 3. The time distribution of $\gamma$-rays for excited states in the daughter populated via the low-spin state $\beta$-decay were added together to create the lower spectrum in Fig.~\ref{fig:decays_half_lives}. The $\gamma$-rays used were the 1087.8, 1113.6, 1149.5, 1188.3, 1208.9, 1226 (triplet), 1240 (doublet), 1276.4, 1303 (triplet), 1351.6, 1683.4, 1811.7, 1856.9, 2277.7, 2287.6, 2333.3, 2357.8, 2389.6, 2396.2, 2435.5, 2464.8, and 2471.8-keV transitions. The doublet and triplet peaks contain multiple photopeaks in them, but all transitions within the multiplets are associated with the low-spin decay of $^{160}$Eu. The extracted half-life from these $\gamma$-rays was $t_{1/2}=26.0(8)$ s. This is approximately 5 seconds shorter than the half-life of $t_{1/2}=30.8(5)$ s from Ref. \cite{Hartley_PRL}, which used the transitions at 1276.4, 1351.6, 2277.7, 2333.3, and 2464.8 keV.

While our analysis used more $\gamma$-rays to determine the half-life in Fig.~\ref{fig:decays_half_lives} than were used in Ref.~\cite{Hartley_PRL}, we also measured the half-life using only the five $\gamma$-rays used in Hartley \textit{et al.}~\cite{Hartley_PRL}. The results obtained via this method were consistent with our measured half-life of $t_{1/2}=26.0(8)$ s for this $\beta$-decay. A ``chop analysis" was performed by systematically varying the first and last channel of the time distribution spectra used for the half-life fit, but no significant dependence on the range of the fit was found in the extracted half-life. Thus, we have no explanation for the discrepancy in the measured half-lives for this $\beta$-decay. However, the agreement between our measured half-life and that of Hartley $et\;al.$ \cite{Hartley_PRL} for the high-spin decay suggests that the discrepancy for this half-life is not due to systematic effects. 

One may wonder if the half-life of the low-spin $\beta$-decay was impacted by contamination of $^{160}$Sm ($t_{1/2}=10.1(11)$ s \cite{nishimura_160Sm_t12}) which populates $^{160}$Eu. 
Overlap of one of the transitions listed above with a strong $\beta$-delayed $\gamma$-ray in $^{160}$Eu of the same energy may shorten the half-life measured. An analysis of the time distribution of individual $\gamma$-rays did not indicate that any of the previously-listed $\gamma$-rays have a significantly-shorter half-life and thus are doublets with transitions in $^{160}$Eu. Furthermore, the use of an element-selective laser ionization scheme would not incidentally ionize Sm, and the use of the IG-LIS source significantly suppressed any surface-ionized species from the ISAC target, including Sm. Thus, we conclude that that the shorter half-life measured here is not due to contaminant $^{160}$Sm in the beam.

\subsection{Lifetimes of Excited States in $^{160}$Gd\label{sec:lifetimes}}
Using the LaBr$_3$(Ce) detectors and the ZDS fast plastic scintillator, we were able to perform first measurements of the lifetime of a number of excited states which are populated in the decay of the high-spin state. Prior to this work, lifetime measurements in $^{160}$Gd have only been completed using Doppler-shift methods, which are generally only sensitive to lifetimes in the sub-picosecond range; few lifetimes in the picosecond to nanosecond range have been reported. 

Our measurements of excited states use two different techniques, both of which are sensitive to lifetimes in the tens of picoseconds or longer range: $\gamma$-$\gamma$ fast-timing using both the convolution fit method and the General Centroid Difference Method (GCDM), and the $\beta$-$\gamma$-$\gamma$ Advanced Time Delayed (ATD) technique using the sequential and parallel transition methods. The convolution fit method and GCDM are both described in Ref.~\cite{GRIFFIN_NIM}, and an example of them applied to GRIFFIN data can be found in Ref.~\cite{Bruno_Hg}. The ATD methods are described in Ref.~\cite{ATD_timing}.

A summary of the lifetimes measured via this analysis is shown in Table~\ref{tab:lifetime_results}.

\subsubsection{1070.8-keV state}
The 1070.8-keV state exhibits a lifetime of the order of nanoseconds; thus, the lifetime of this state can be extracted via a fit of TAC spectra with the convolution of a Gaussian and exponential decay. Because this state has two strong depopulating transitions -- the 822.2- and 995.5-keV $\gamma$-rays -- the lifetime can be measured via both, the 412.6 -- 822.2 and 412.6 -- 995.5-keV cascades.

The 412.6 -- 822.2-keV cascade resulted in a lifetime of $\tau=1620(180)$ ps, while the 412.6 -- 995.5 keV cascade gave a lifetime of $\tau=1630(180)$ ps. We adopt a lifetime of $\tau=1.63(13)$ ns for this state, which is the average of the two measurements.

\begin{figure}[!htb]
    \includegraphics[width=1\linewidth]{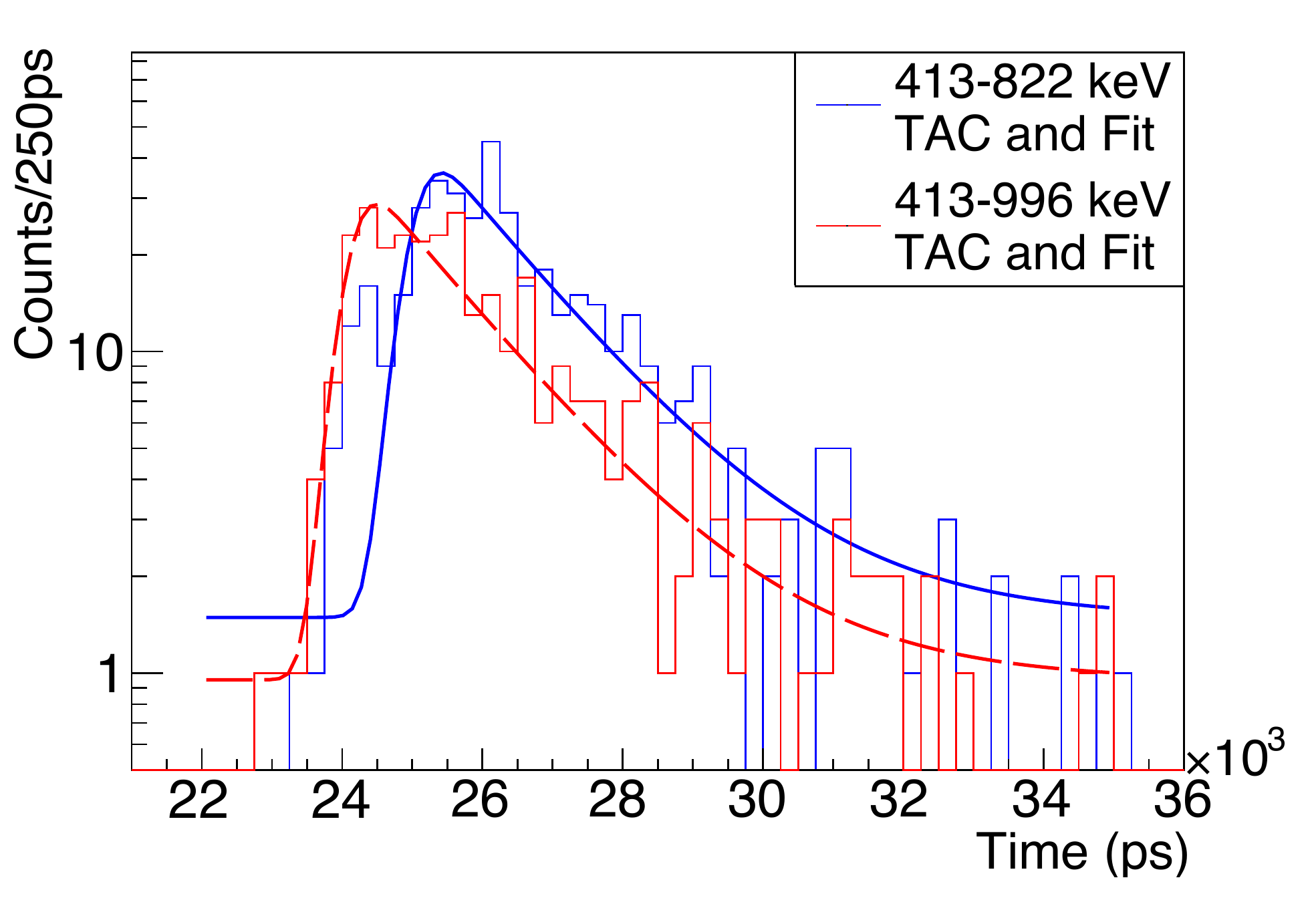}
    \caption{(Color online) TAC spectra for the 1070.8-keV state. The 412.6 -- 822.2-keV cascade and fit are shown in blue, and the 412.6 -- 995.5-keV cascade and fit in red. The adopted lifetime for this state from these two fits is $\tau=1630(130)$ ps. 
    }
    \label{fig:1071_TACs}
\end{figure}

\subsubsection{1483.4-keV state}
\begin{figure}[!htb]
    \includegraphics[width=1\linewidth]{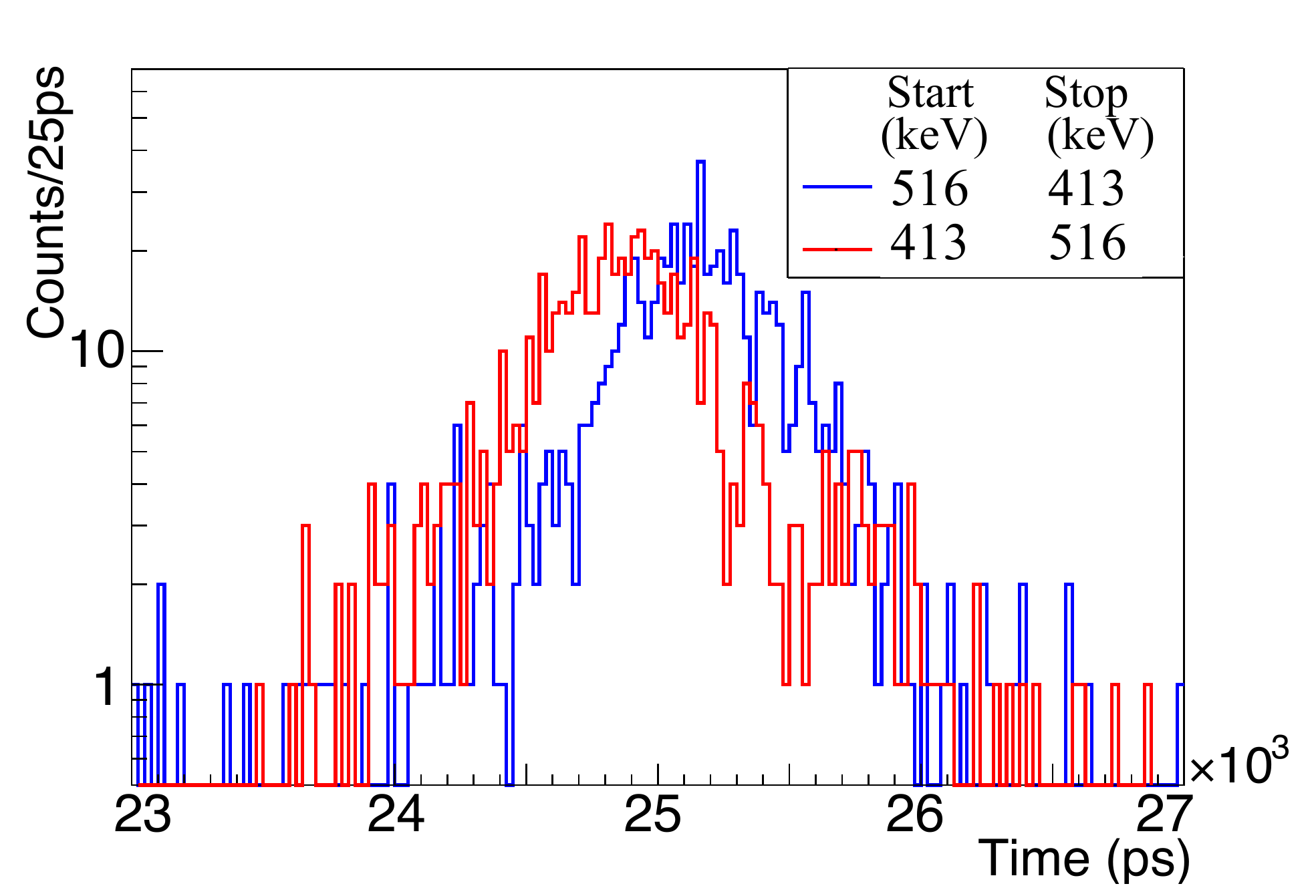}
    \caption{(Color online) TAC spectra for the 1483.4-keV state using $\gamma$-$\gamma$ timing. The normal gate with the TAC started by the populating 515.6-keV transition and stopped by the depopulating 412.6-keV transition is shown in blue; the antigate (412.6-keV transition as start and 515.6-keV as stop) is shown in red. The resulting lifetime is $\tau=97(22)$ ps. \iffalse Note that the centroid location of each TAC has been corrected to account for the Compton background contribution to the centroid.\fi}
    \label{fig:1483_TAC}
\end{figure}

The lifetime of the 1483.4-keV state can be measured via the intense 515.6 -- 412.6-keV cascade. Because the lifetime of the 1483.4-keV state is sufficiently short, we were required to use the GCDM technique to measure its lifetime. After determining the centroid difference ($\Delta C=218(34)$ ps) and corrections for the time-walk of the setup (PRD$=23(28)$ ps), we measured a lifetime of $\tau=97(22)$ ps for the state. The associated TAC spectrum is shown in Fig.~\ref{fig:1483_TAC}.

\subsubsection{1581.9-keV state}

The lifetime of the 1581.9-keV state was measured with $\beta$-$\gamma$-$\gamma$ timing using the sequential transition method. HPGe and LaBr$_3$(Ce) gates were placed on the 417.0- and 408.5-keV transitions, and the centroid locations for each TAC spectra were subtracted to determine the lifetime of the 1581.9 keV state. In this analysis, we have assumed that the prompt time response of the 408.5-keV and 417.0-keV $\gamma$-rays in the LaBr$_3$(Ce) detectors was the same due to their similar energy.

\begin{figure}
    \centering
    \includegraphics[width=1\textwidth]{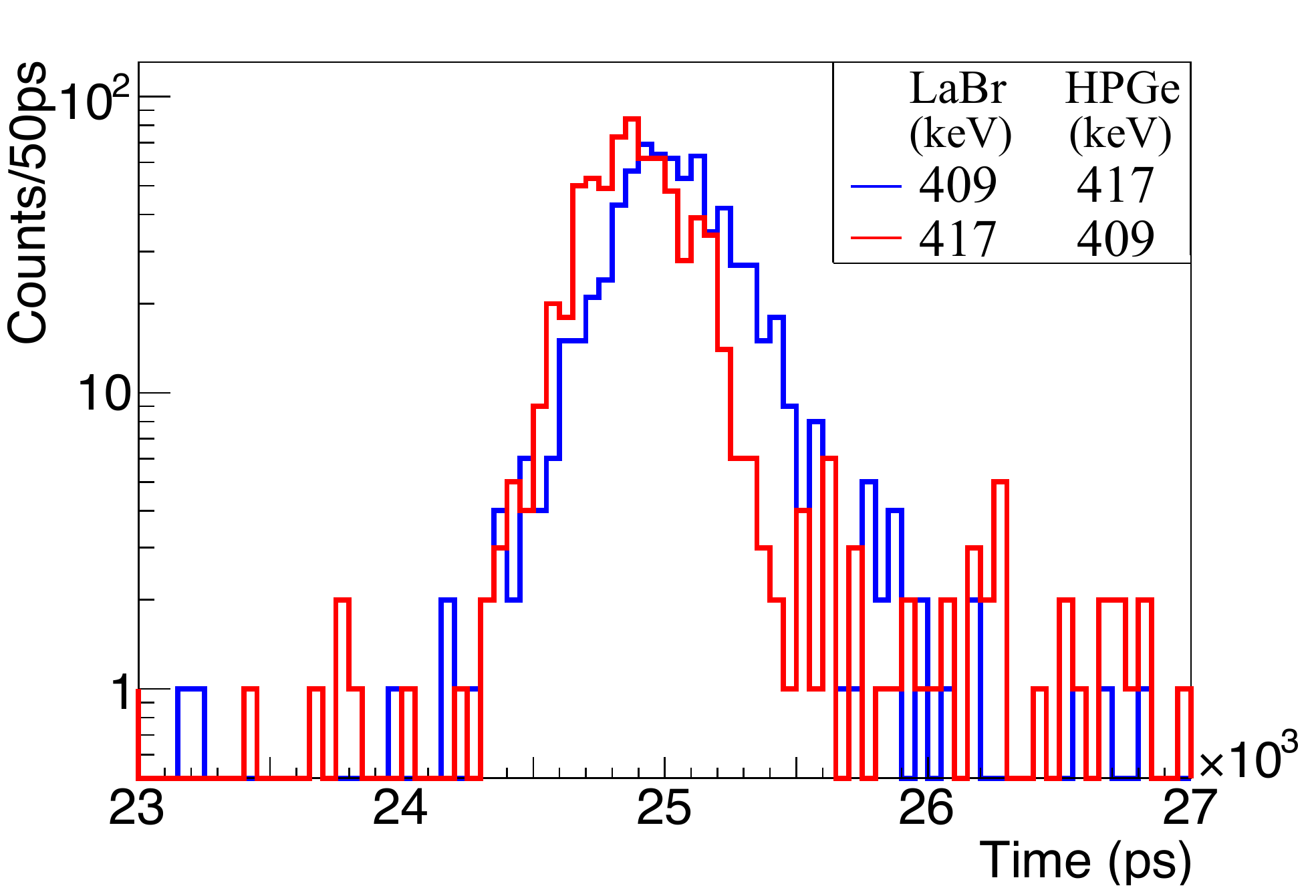}
    \caption{(Color online) TAC spectrum for the 1581.9-keV state. The TAC is started via the $\beta$-particle and stopped in the LaBr$_3$(Ce) detector, with the HPGe gate used to select the cascade. From this, a lifetime for the 1581.9-keV state of $\tau=160(20)$ ps is deduced.}
    \label{fig:1582_beta_gamma}
\end{figure}

After corrections for the Compton background contribution, the lifetime of the 1581.9-keV state was measured to be $\tau=160(20)$ ps. The TAC spectra for each gate is shown in Fig.~\ref{fig:1582_beta_gamma}.

A consequence of this lifetime measurement is a constraint on the mixing ratio of the 98.6-keV transition. If the 98.6-keV transition were pure $E2$ multipolarity, it would have a reduced transition rate of B($E2$)$\approx$2100 W.u., substantially above any reasonable value in this mass region. Taking an upper-bound of B($E2$)$<$300 W.u. for the 98.6-keV transition gives $|\delta|<0.41$ for this transition.

\subsubsection{1999.0-keV state}
Because the 1999.0-keV state is heavily populated via $\beta$-decay, there are insufficient statistics available to measure the lifetime directly via $\gamma$-$\gamma$ fast-timing in this analysis. However, we were able to set an upper limit on the lifetime of this state using $\beta$-$\gamma$-$\gamma$ timing. 

We utilized the $\beta$-$\gamma$-$\gamma$ parallel transition method originating from both the 2490.2- and 2560.8-keV states. The TAC is started by detection of the $\beta$-particle and stopped with the detection of the 412.6-keV $\gamma$-ray in the LaBr$_3$(Ce) detectors. The HPGe gate is used to select a cascade that runs through the 1999.0-keV state and a cascade that bypasses it. Because both cascades start at the same level and stop with the same $\gamma$-ray in the LaBr$_3$(Ce) detectors, the prompt time response of the LaBr$_3$(Ce) detectors is the same in both cases. Subtracting the centroid locations from each then directly gives the lifetime of the intermediate 1999.0-keV state.

\begin{figure}
    \centering
    \includegraphics[width=1\textwidth]{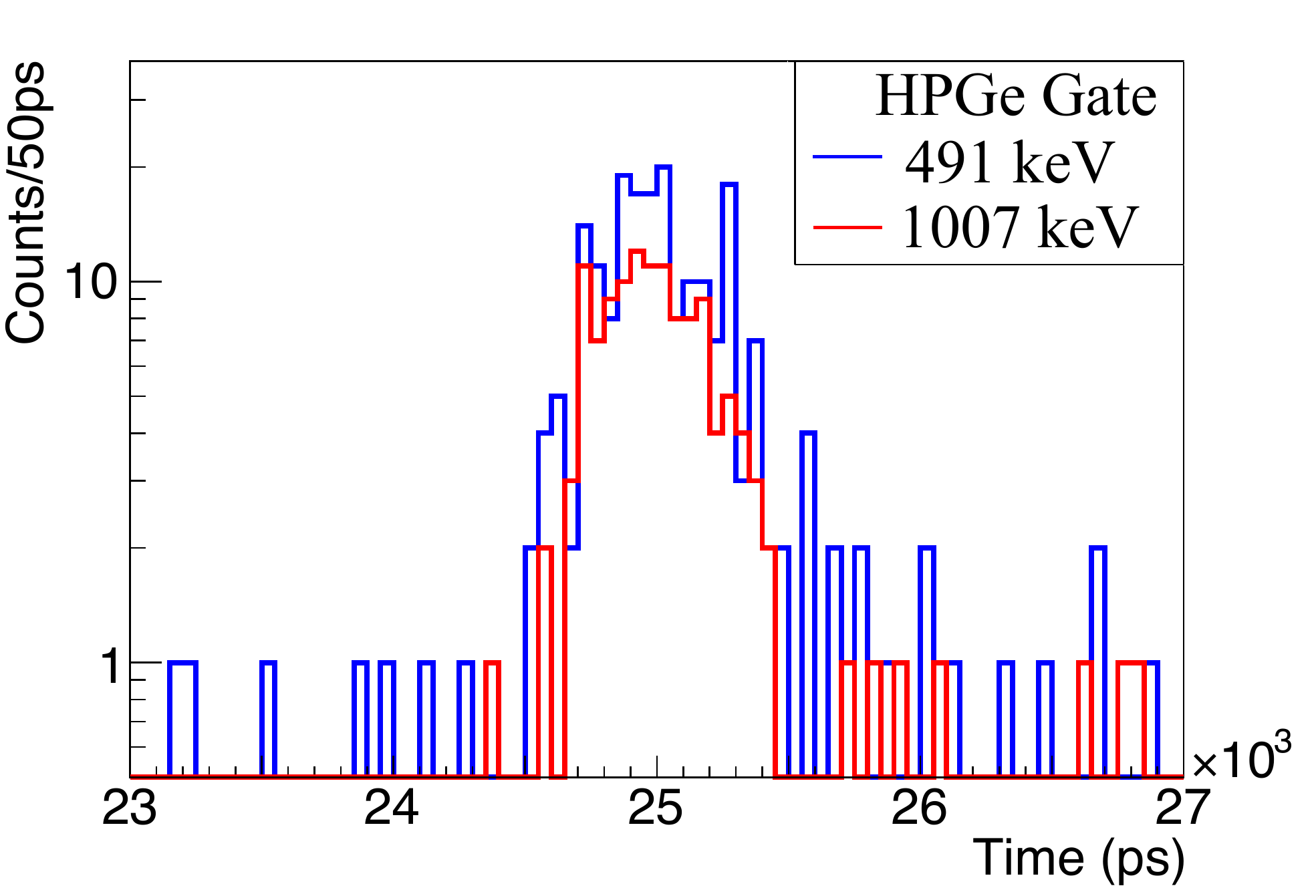}
    \caption{(Color online) TAC spectrum for the 1999.0-keV state using the $\beta$-$\gamma$ parallel transitions timing method. The TAC is started by the $\beta$-particle and stopped by a detection of a 412.6-keV $\gamma$-ray in the LaBr$_3$(Ce) detector. The HPGe gate is used to select the cascade of interest. We see no significant shift from prompt of the TAC spectra, indicating an upper limit on the lifetime of $\tau<100$ ps.}
    \label{fig:1999_lifetime_beta_gamma}
\end{figure}

The parallel cascade pair for the 2490.2-keV state are the 490.7- and 1006.8-keV $\gamma$-rays, and for the 2560.8-keV state the $\gamma$-rays are 560.8- and 1076.8-keV. In both cases, no significant lifetime is seen for the 1999.0-keV state. Based on our timing sensitivity, we set a conservative upper limit on the lifetime of the 1999.0-keV state at $\tau<100$ ps. The TAC spectra for the parallel method using the $\gamma$-rays from the 2490.2-keV state are shown in Fig.~\ref{fig:1999_lifetime_beta_gamma}.

\begin{longtable}{c|c|c}
    \caption{Measured lifetimes in $^{160}$Gd from $\gamma$-$\gamma$ and $\beta$-$\gamma$-$\gamma$ fast-timing analysis.}\\
    \hline\hline
     State Energy (keV) &  Lifetime (ps)& Method  \\ \hline
     \endfirsthead
     \endhead
     \hline\hline
     \endlastfoot
   
     1070.8 & 1630(130) & $\gamma$-$\gamma$ \\
     1483.4 & 97(22)    & $\gamma$-$\gamma$ \\
     1581.9 & 160(20)   & $\beta$-$\gamma$-$\gamma$ \\
     1999.0 & $<$100   & $\beta$-$\gamma$-$\gamma$

     \label{tab:lifetime_results}
     
\end{longtable}


\subsection{$\gamma$-$\gamma$ Angular Correlations\label{sec:ggAC}}

$\gamma$-$\gamma$ angular correlations were performed on the most intense cascades present in $^{160}$Gd using the ``Method 4" procedure outlined in Ref.~\cite{GRIFFIN_NIM_ANG_CORR} which uses pre-calculated attenuation coefficients for the angular correlation $Z$-distributions. The normalized counts vs. cos($\theta$) graph is fitted with a linear combination of even Legendre polynomials and the angular correlation coefficients extracted. Normalization of the counts for each angle was performed using the ``event-mixing" technique outlined in Ref.~\cite{GRIFFIN_NIM_ANG_CORR}. 

To verify the method for this experiment, the $\gamma$-$\gamma$ angular correlation analysis was performed on data from the 778.9 -- 344.3-keV $3^-\rightarrow 2^+\rightarrow 0^+$ cascade in $^{152}$Gd taken from a $^{152}$Eu source run used for post-experiment calibrations. The extracted $a_2$ and $a_4$ angular correlation parameters for this cascade were consistent with literature values.

A summary of the $\gamma$-$\gamma$ angular correlation fit parameters and resulting transition mixing ratios is shown in Table~\ref{tab:AC_parameters}. Reduced transition rates for transitions from the 1070.8, 1483.4, 1581.9, and 1999.0 keV states based on measured lifetimes and mixing ratios are shown in Table~\ref{tab:red_rates_sum}.

\subsubsection{822.2 -- 173.2-keV Cascade}

\begin{figure}
    \centering
    \includegraphics[width=0.9\linewidth]{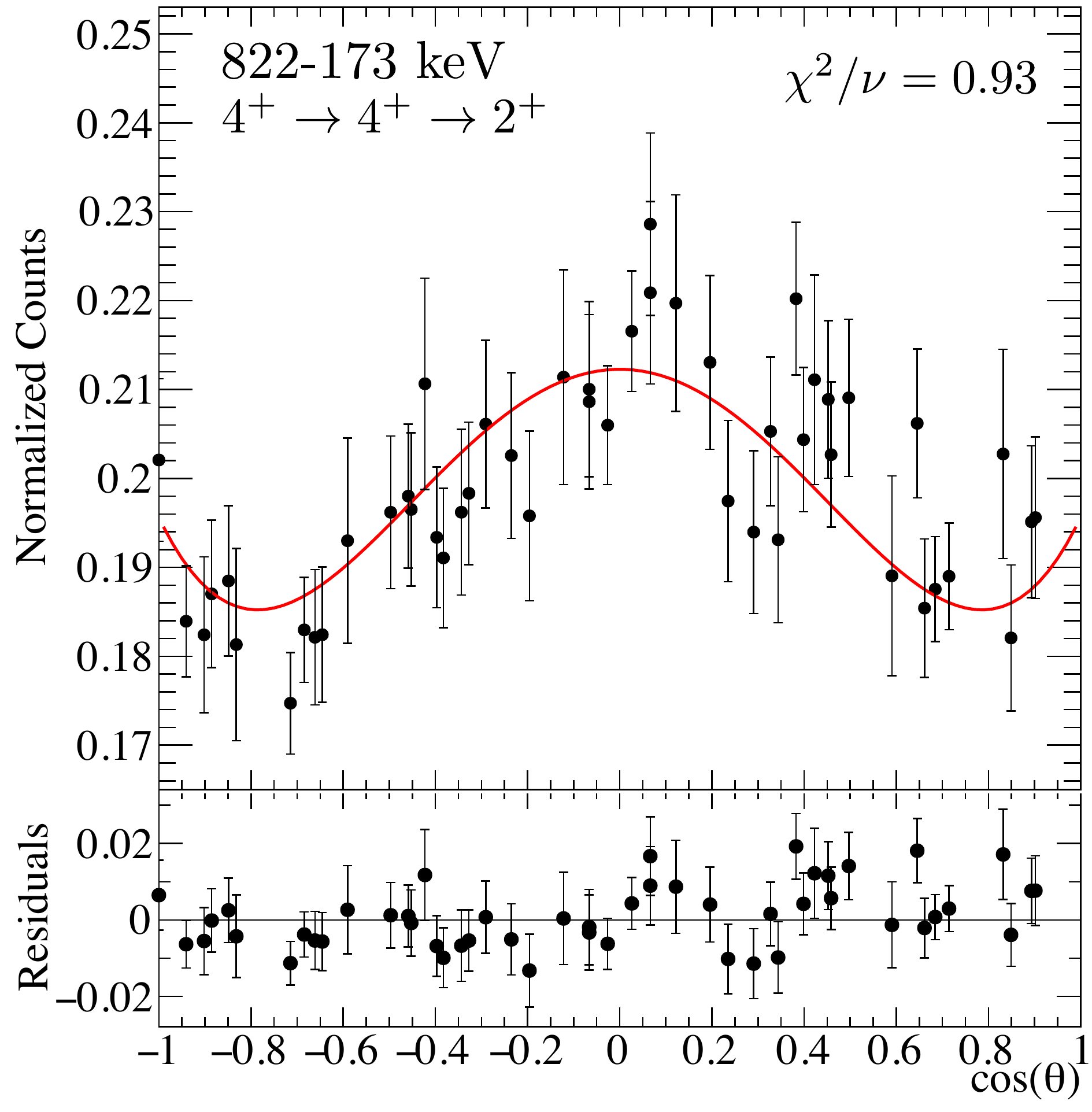}
    \caption{Angular correlation for the 822.2 -- 173.2-keV cascade. The best-fit parameters are $a_2=-0.10(1)$ and $a_4=0.10(2)$ with a reduced chi-squared for the fit of $\chi^2/\nu=0.93$.}
    \label{fig:ac_822-173}
\end{figure}

Angular correlations were performed for the 822.2 -- 173.2 keV cascade with the initial $\gamma$-ray depopulating the 1070.8-keV state. The resulting angular correlation graph is shown in Fig.~\ref{fig:ac_822-173}. With $\chi^2/\nu=0.93$, the best-fit results indicate angular correlation coefficients of $a_2=-0.10(1)$ and $a_4=0.10(2)$ for this $4^+\rightarrow 4^+\rightarrow 2^+$ cascade.

\begin{figure}
    \centering
    \includegraphics[width=1\linewidth]{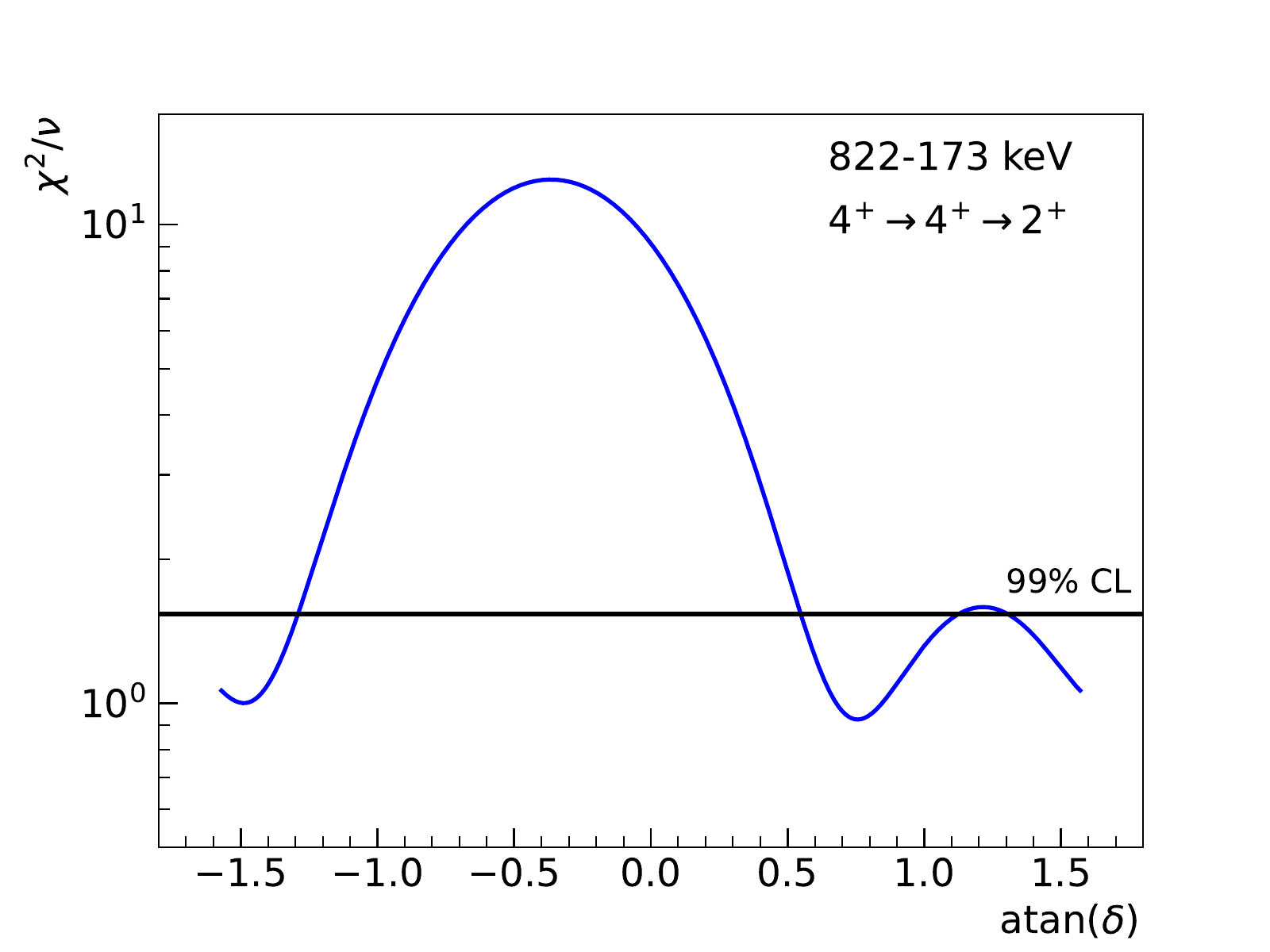}
    \caption{(Color online) $\chi^2/\nu$ versus atan($\delta$) for the 822.2-keV $\gamma$-ray using the 822.2 -- 173.2-keV $4^+\rightarrow4^+\rightarrow2^+$ cascade. The mixing ratio of the 173.2-keV $\gamma$-ray was fixed at $\delta=0$. See text for more details.}
    \label{fig:MR_822_173}
\end{figure}

Based on the angular correlations, the possible mixing ratios for the 822.2-keV transition ($\delta$) are shown in Fig.~\ref{fig:MR_822_173}. The $4^+\rightarrow 2^+$ 173.2-keV transition's mixing ratio was fixed at $\delta=0$ as it can be assumed to be pure $E2$. From the analysis, the most likely mixing ratios for the 822.2-keV transition are $\delta=0.94^{+9}_{-8}$ with $\chi^2/\nu=0.92$ and $\delta=-8.1^{+2}_{-183}$ with $\chi^2/\nu=1.00$. The uncertainty on the mixing ratio corresponds to the values of the mixing ratio minimization associated with $\Delta\chi^2=+1$.

These values contrast with the mixing ratio measured in Ref.~\cite{Govor} of $\delta_{822}=-0.71(3)$. That measurement was done via the ($n,n'\gamma$) inelastic scattering reaction and used used $\gamma$-ray singles spectra collected at seven angles to determine the $\gamma$-ray angular distribution. The data presented here, in contrast, arise from $\gamma$--$\gamma$ coincidences. Additionally, GRIFFIN uses 52 angular data points for the angular correlation. Thus, we deem the current analysis to provide a more reliable result that measured in Ref.~\cite{Govor}.


\subsubsection{515.6 -- 412.6-keV Cascade}

\begin{figure}[!h!]
    \centering
    \includegraphics[width=0.9\linewidth]{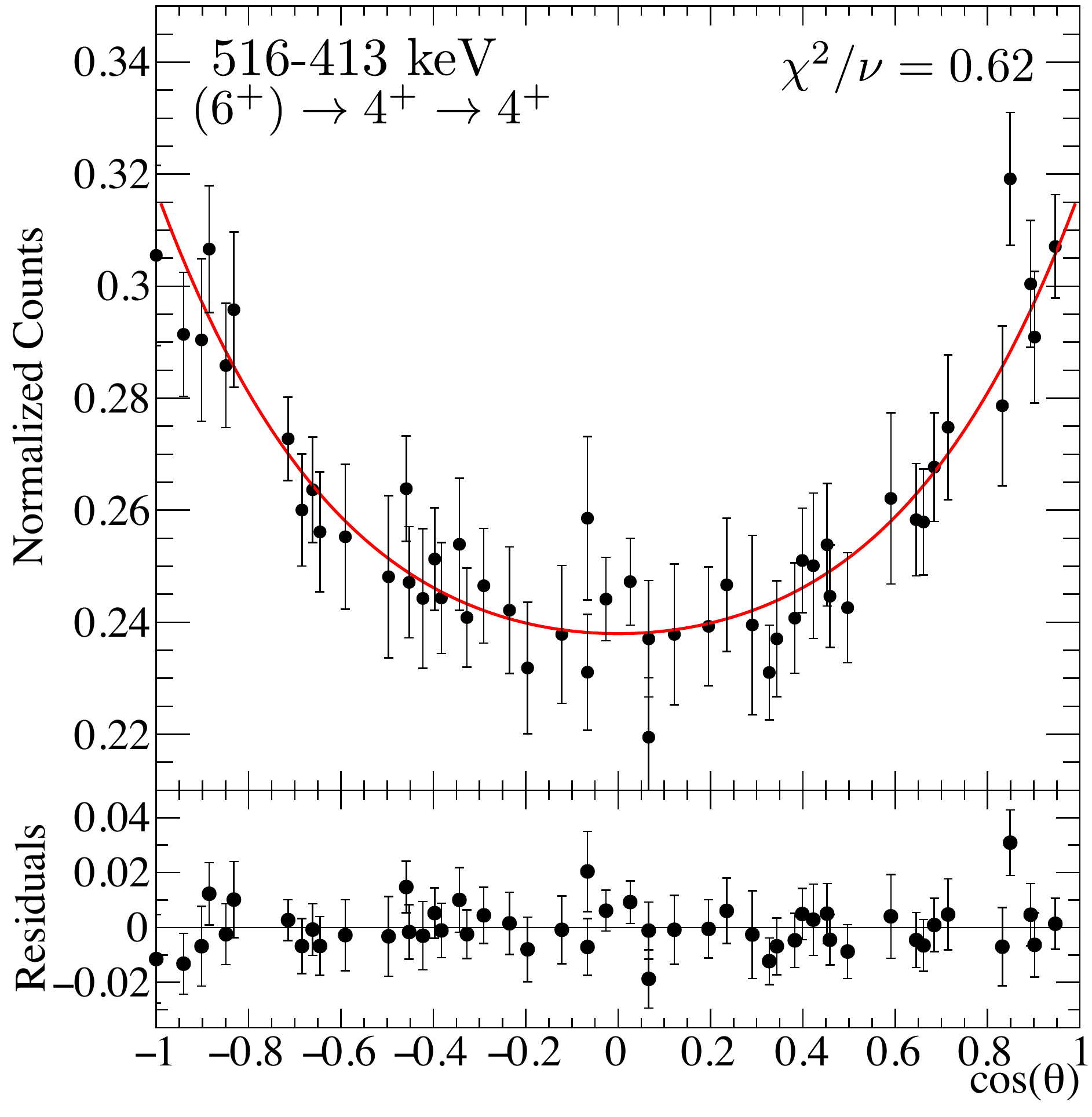}
    \caption{Angular correlation for the 515.6 -- 412.6-keV cascade. The best-fit parameters are $a_2=0.20(1)$ and $a_4=0.03(2)$ with a reduced $\chi^2$ for the fit of $\chi^2/\nu=0.62$.}
    \label{fig:ac_516_413}
\end{figure}

Figure~\ref{fig:ac_516_413} shows the angular correlation plot for the 515.6 -- 412.6-keV cascade originating from the 1999.0-keV state. The best-fit correlation gives values of $a_2=0.20(1)$, $a_4=0.03(2)$, and $\chi^2/\nu=0.62$.

\begin{figure}
    \centering
    \includegraphics[width=1\columnwidth]{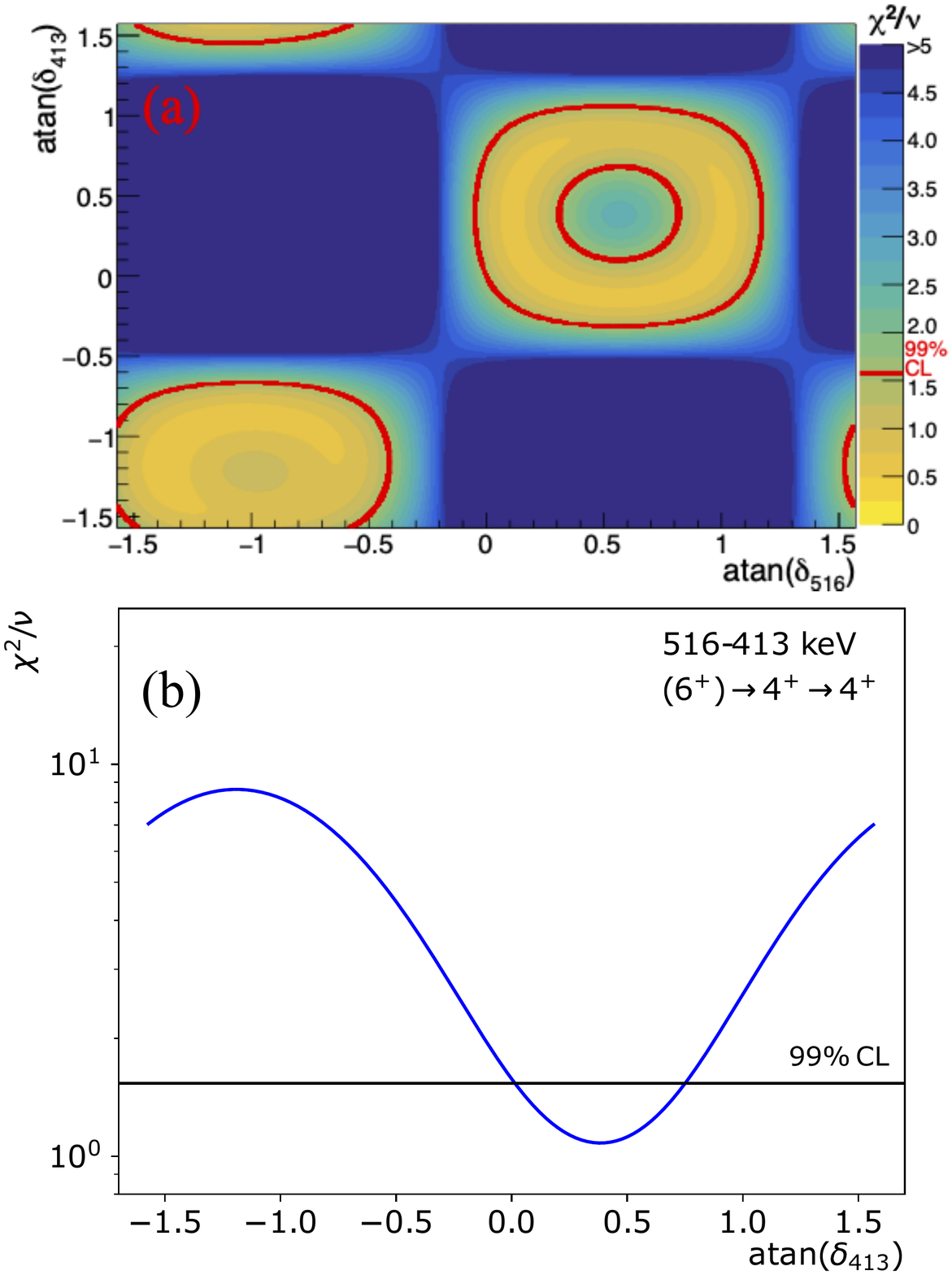}
    \caption{(Color online) (a) Mixing ratio combinations for the 515.6- and 412.6-keV transitions based on the 515.6 -- 412.6-keV $(6^+)\rightarrow4^+\rightarrow4^+$ angular correlation. The red contours denote mixing ratio combinations where the reduced $\chi^2$ is at the 99\% confidence level (CL), with regions in yellow indicating mixing ratio combinations above the 99\% CL. (b) $\chi^2/\nu$ versus atan($\delta$) for the 412.6-keV $\gamma$-ray based on a mixing ratio of the 515.6-keV transition of $\delta=0$ (see text for details). The minimized mixing ratio occurs at $\delta=0.41(10)$.}
    \label{fig:MR_516_413}
\end{figure}

Fig.~\ref{fig:MR_516_413} shows the possible mixing ratio combinations for the 515.6-keV ($\delta_{516}$) and 412.6-keV ($\delta_{413}$) transitions based on the 515.6 -- 412.6-keV cascade and under the assumption that the spin of the 1999.0 keV state is $J^\pi=6^+$, which is discussed in detail in Sec.~\ref{Sec:Disc}. The red contours in Fig.~\ref{fig:MR_516_413} denote $\chi^2/\nu$ values at the 99\% confidence level (CL), with yellow regions having $\chi^2/\nu$ values above the 99\% CL. Four regions of possible mixing ratio combinations appear in Fig.~\ref{fig:MR_516_413}(a).

The angular correlation alone cannot constrain the mixing ratios of these transitions. However, based on the measured lifetime of the 1999.0-keV state of $\tau<100$ ps (Sec.~\ref{sec:lifetimes}), the mixing ratio of the 515.6-keV transition can be constrained to $|\delta|<3.6\times10^{-4}$. A detailed discussion of this constraint can be found in Sec.~\ref{sec:1999_state}. With this constraint, a 1-D minimization of the mixing ratio of the 412.6-keV transition (Fig.~\ref{fig:MR_516_413}(b)) yields $\delta=0.41(10)$.


\begin{longtable}{c|cc|c|c}
    \caption{$\gamma$-$\gamma$ angular correlation parameters and associated best-fit mixing ratios for intense cascades in $^{160}$Gd. }\\
    \hline\hline
     $\gamma_1$-$\gamma_2$ &  $a_2$ & $a_4$ & $\delta_1$ & $\delta_2$ \\ \hline
     \endfirsthead
     \endhead
     \hline\hline
     \caption*{\footnotesize{$^a\;\delta_\mathrm{lit}=-0.71(3)$ \cite{Govor}.\\\hspace{\textwidth}$^b$\;$\delta_2$ fixed at 0.}}
     \endlastfoot
     822-173  & -0.10(1) & 0.10(2) & $-8.1^{+2}_{-183}$ $^a$   & 0 $^b$ \\
         $4^+$-$4^+$-$2^+$    &           &        & $0.94^{+9}_{-8}$ $^a$ & 0 $^b$ \\
         \hline
        516-413 & 0.20(1) & 0.03(2) & $|\delta_{1}|<3.6\times10^{-4}$  & $0.41(10)$  \\
             $(6^+)$-$4^+$-$4^+$   &         &         & & 
        \label{tab:AC_parameters}
     
\end{longtable}


\renewcommand{\arraystretch}{1.5}


\section{Discussion}\label{Sec:Disc}


\subsection{1999.0-keV State Configuration\label{sec:1999_state}}

The dominant high-spin isomeric $\beta$-decay of $^{160}$Eu proceeds to the state at 1999.0 keV.
Ref.~\cite{Hartley_PRL} assigns this state as $5^-$ with a configuration of $\pi^2(5/2[413],5/2[532])$ based on the proposed $\pi 5/2[413] \otimes \nu 5/2[523]$ configuration for the $5^-$ ground-state of the parent $^{160}$Eu, the observed log($ft$) value of 5.1, and deformed Woods-Saxon calculations.

It is interesting to note, as also discussed in Ref.~\cite{Hartley_PRC}, that based on the above assignment, one would expect an allowed decay to a $K^{\pi} = 6^-$ state, based on the $\pi^2(5/2[413],7/2[523])$ configuration, with a log($ft$) value of $\sim$4.8 for the  $\nu 5/2[523] \rightarrow \pi 7/2[523]$ transition \cite{Sood,bohr_mottelson_ii}, for which no candidates have been proposed. 

The observation of the 450.4-keV transition from the 1999.0-keV state to the 1548.5-keV $7^+$ member of the $\gamma$-band is curious as already pointed out in Ref.~\cite{Hartley_PRC}. If the 1999.0 keV state is indeed a $5^{-}$ state, this transition would be of $M2$ nature. Hartley \textit{et al.} stated that no measurable extended lifetime was observed for the 1999.0-keV state but did not quote an upper limit. Also at the time of that publication, the $7^+$ assignment for the 1548.5-keV state was tentative, but in their recent publication using Coulomb excitation of $^{160}$Gd \cite{Hartley_gd_coulex}, this state was firmly established as a member of the $\gamma$-band.

Our analysis set an upper limit of $\tau<100$ ps on the lifetime of the 1999.0-keV state based on $\beta$-$\gamma$-$\gamma$ fast-timing as discussed in Sec.~\ref{sec:lifetimes}. This contrasts with the expected long lifetime described in Ref.~\cite{Hartley_PRC} and casts doubt on the $5^-$ assignment for the 1999.0-keV state. Based on the $\gamma$-decay patterns from this state, it is expected to have spin  $4^\pm$, $5^\pm$ or $6^\pm$.  We examine each possibility below:

A $4^\pm$ assignment can be ruled out based on the decay to the $7^+$ state, as that would require an $L=3$ transition with a reduced transition rate on the order of 1000's of W.u. A $5^-$ assignment, as suggested by Hartley \textit{et al.} \cite{Hartley_PRL,Hartley_PRC}, would make the 450.4-keV transition $M2$ in nature with $B(M2)>15$ W.u. Data on $M2$ transitions in this region is scarce, but Ref.~\cite{endt_gamma_strengths} notes that in the $A=91-150$ region, no $M2$ transitions of strength greater than 1 W.u. have been observed. Alternatively, a $6^-$ assignment similarly makes the intense 515.6-keV transition $M2$ in nature, with a pure $M2$ transition strength of $B(M2)>280$ W.u. One may wonder if these transitions may be mixed $M2/E3$; however, a mixing ratio that brings the $M2$ component in line with observed values ($<1$ W.u.) then leads to an $E3$ component with strength of 1000's of W.u. From this, the $4^\pm$, $5^-$, and $6^-$ spin assignments are all eliminated for the 1999.0-keV state.

In contrast, for a spin assignment of $5^+$ or $6^+$, all transitions from the 1999.0-keV state would be $M1$ and $E2$ in character. Reduced transition rates for these transitions depend on the spin assignment and mixing ratio of the transitions, but they are on the order of $B(M1)\approx10^{-5}-10^{-3}$ W.u. and $B(E2)\approx0.1-10$ W.u. These match well with known transitions rates in the nearby $A=91-150$ mass region \cite{endt_gamma_strengths}.

Thus, based purely on the experimental lifetime limit and observed transitions to states with well established spin and parity assignments, the 1999.0 keV is most likely a $5^+$ or $6^+$ state.

Further insight on the most plausible spin and parity for the 1999.0-keV state can be gained by considering the possible Nilsson configurations for $5^+$ or $6^+$ states at this energy. In this region, there are several available two-quasiproton configurations that can create spin $5^+$ and $6^+$. A $5^+$ state can be formed by the $\pi^2(7/2[523],3/2[541])$ configuration. However, the $\pi3/2[541]$ orbital lies well below the Fermi surface at the known (ground-state) deformation $\beta_2\approx0.35$ for this nucleus \cite{PRITYCHENKO_adndt_2016}. Access to the $\pi3/2[541]$ orbital would require the deformation of this excited state to be significantly lower ($\beta_2\approx0.1$) than the known deformation of the ground-state of $^{160}$Gd. A $6^+$ state can be formed by the $\pi^2(5/2[413], 7/2[404] )$ and $\pi^2(7/2[523],5/2[532])$ configurations. The  $\pi^2(5/2[413], 7/2[404)$ $6^+$ state is calculated to be at 2.5 MeV \cite{Hartley_PRC} but cannot be populated by an allowed-unhindered $\beta$-decay from $^{160}$Eu. This leaves the  $\pi^2(7/2[523],5/2[532])$ $6^+$ state. Recent self-consistent Deformed Hartree-Fock and Angular Momentum Projection calculations for heavier Gd isotopes \cite{Ghorui_2018} predict the $\pi^2(7/2[523],5/2[532])$ $6^+$ state in $^{164,166,168}$Gd at $\sim$2 MeV, which makes it plausible to assume that the 1999.0 keV state in $^{160}$Gd might indeed be a $6^+$ state associated with the $\pi^2(7/2[523],5/2[532])$ configuration.

One consequence of this spin assignment is a significant constraint on the mixing ratio of the 515.6-keV transition (and thus the 412.6-keV transition that follows it). With the $J^\pi$=$(6^+)$ spin assignment for the 1999.0-keV state, the 515.6-keV transition depopulating it is $E2/M3$ in character. Based on the lifetime limit of the 1999.0-keV state at $\tau<100$ ps (see Sec.~\ref{sec:lifetimes}), the 515.6-keV transition must be essentially pure $E2$. As previously mentioned, $\gamma$-ray strength in this region are not well documented, but in the $A=91-150$ region, no $B(M3)$ values above 10 W.u. have been observed \cite{endt_gamma_strengths}. In order for the $M3$ component of the 515.6-keV transition to have $B(M3)<10$ W.u., the mixing ratio of the transition must be $|\delta|<3.6\times10^{-4}$, which gives a $M3$ branching ratio of $\sim10^{-7}$; hereafter assumed to pure $E2$ transition.

With the mixing ratio of the 515.6 keV $\gamma$-ray fixed at $\delta=0$, a 
minimization of the second $\gamma$-ray in the cascade (412.6 keV) can occur. The minimized mixing ratio plot is shown in Fig.~\ref{fig:MR_516_413}, with a best-fit mixing ratio for the 412.6 keV transition occurring at $\delta=0.41(10)$. This minimized value is independent of the choice of mixing ratio for the 515.6 keV transition for $|\delta|<3.6\times10^{-4}$.

This understanding of the Nilsson orbital structure of the 1999.0-keV state is an important observation in the context of the discussion of the two $K^\pi$=$4^+$, $4^+$ states at 1070.8 keV and 1483.4 keV. If, as discussed in more detail later in this section, these two states have very similar quasiparticle configurations and the decay of the 1999.0-keV state to the 1483.4 keV $4^+$ state were an allowed $E1$ transition between the $\pi 5/2[532]$ and $\pi 3/2[411]$ orbitals, as suggested in Ref.~\cite{Hartley_PRL,Hartley_PRC}, one naturally would expect to observe a decay of the 1999.0-keV state to the 1070.8-keV $4^+$ state that is favored over the decay to the 1483.4-keV $4^+$ state. However, in this work, no such decay is observed. 
In fact, the transition between the $\pi 5/2[532]$ and $\pi 3/2[411]$ orbitals involves $\Delta n_z=2$ and is forbidden. The non-observation of a transition from the 1999.0-keV $6^+$ state to the 1070.8-keV $4^+$ state may suggest that the transition to the 1483.4-keV $4^+$ state is facilitated through a small wavefunction component that is not in common between the two $K^\pi$=$4^+$ bands.


\subsection{$^{160}$Eu Configuration\label{sec:160Eu_configs}}

Hartley $et\;al.$~\cite{Hartley_PRL,Hartley_PRC} proposed a configuration for the ground-state of $^{160}$Eu to be $\pi 5/2[413] \otimes \nu 5/2[523]$, with the $\beta$-decay to the 1999.0 keV state in $^{160}$Eu understood to be a $\nu5/2[523]\rightarrow \pi5/2[532]$ transition \cite{Hartley_PRL}. 
This is indeed the assignment that one would naturally expect based on systematic trends and firm spin-parity assignments in neighboring odd-A nuclei. However, based on the observed $\beta$-decay intensities, $\gamma$-ray transitions and the knowledge that the 1999.0-keV state in $^{160}$Gd has positive parity, the high-spin $\beta$-decay of $^{160}$Eu is not easily understood. We discuss two possibilities here, neither of which can be conclusively confirmed nor rejected: {\it (i)} the parent state in $^{160}$Eu has negative parity and $\beta$-decays to a negative parity $6^-$ state just above the 1999.0-keV state, or {\it (ii)} the parent is positive parity and undergoes direct $\beta$-decay to the 1999.0-keV state.

Hartley $et\;al.$ \cite{Hartley_PRC} suggested the possible presence of a $6^-$ state just above the 1999.0-keV state that is populated via $\beta$-decay and undergoes a highly-converted transition to the 1999.0-keV state. They did not provide any evidence for or against this possibility, however. We put the hypothesis of a $6^-$ state around 2 MeV to the test. We were unable to find any viable peaks in the conversion electron spectrum for this transition, as all intense peaks in the spectrum are associated with known transitions in the de-excitation of $^{160}$Gd. Our lower bound for conversion electron detection is around 20 keV. However, based on the fact that we do not see an extended lifetime in the $\beta$-$\gamma$-$\gamma$ fast-timing, this theoretical highly-converted $E1$ transition would require $E_\gamma\gtrsim$ 50 keV in order to obtain $B(E1)\lesssim10^{-2}$ W.u. Given that all of the transition intensity from the 1999.0 keV would flow through this transition, the intensity of the conversion electron would be at least half that of the 75.3-keV $2^+_{1}\rightarrow 0^+_{g.s.}$ transition. Based on this information, the L-shell line of this the transition from this hypothetical $6^-$ state to the 1999.0-keV state should have been clearly visible in the conversion electron spectrum; no such transition was observed, however. 

\begin{figure}
    \centering
    \includegraphics[width=1\columnwidth]{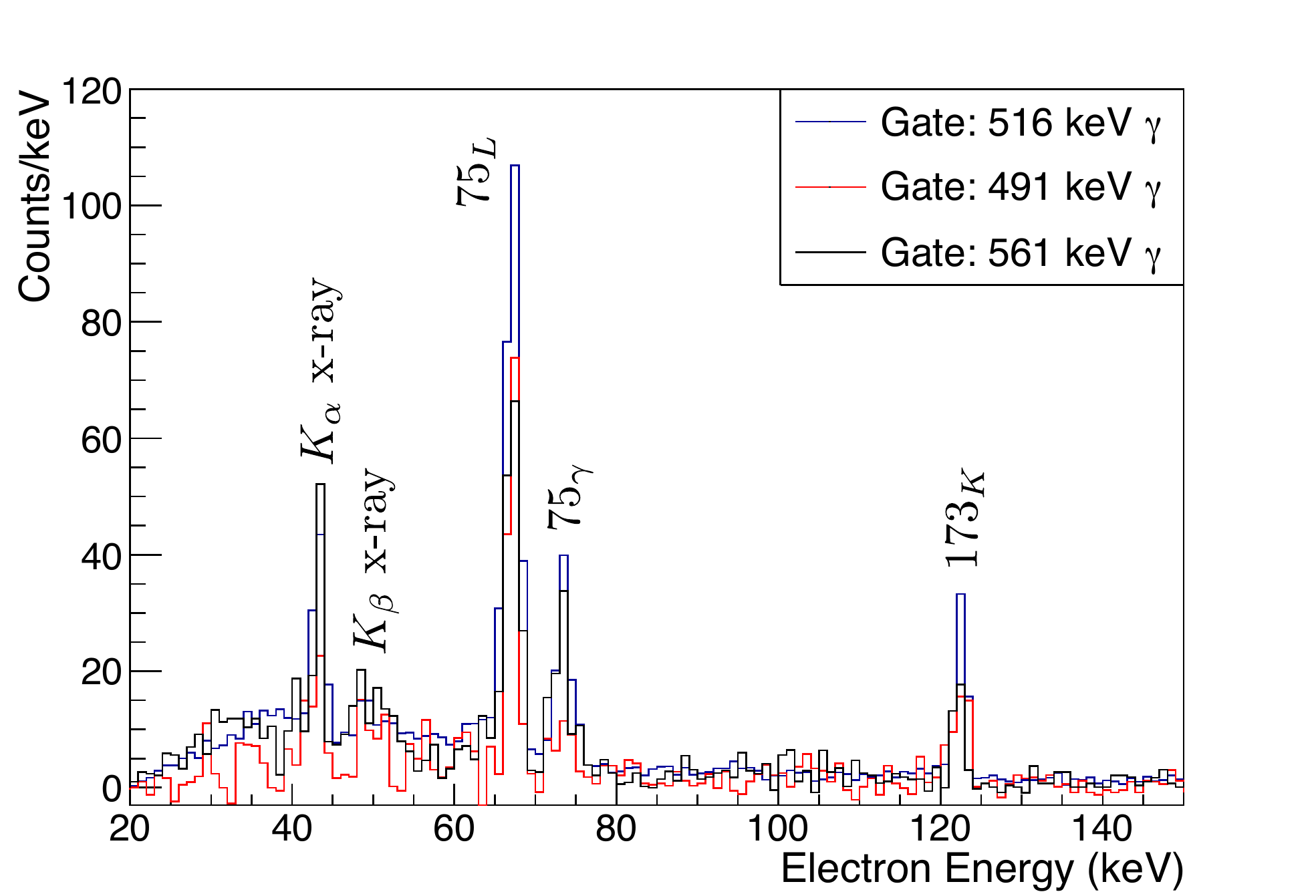}
    \caption{(Color online) Comparison of $\gamma$-gated conversion electron spectra in $^{160}$Gd. The presence of a highly-converted transition from a $6^-$ state just above 1999.0 keV would appear in the 515.6-keV $\gamma$-gated spectrum (blue) and not in the 490.7- (red) and 560.8-(black) keV $\gamma$-gated spectra. No transition is seen. Note that the 515.6-keV $\gamma$-gated spectrum has been downscaled approximately 90\% to match the background levels of the 490.7- and 560.8-keV spectra. See text for more details.}
    \label{fig:CE_comparison_516_560}
\end{figure}

Additionally, we would have expected to observe the conversion electrons of this transition in coincidence with the 515.6-keV transition that depopulates the 1999.0-keV state. At the same time, this transition would not be seen in coincidence with the 490.7- and 560.8-keV transitions that populate the 1999.0-keV state. By comparing the $\gamma$-ray gated conversion electron spectra, this highly-converted $E1$ transition should have been visible. We do not see evidence of a conversion electron peak in the 515.6 keV gated spectrum that is not present in the 490.7- and 560.8-keV gated spectra. The comparison of the conversion electron spectra are shown in Fig.~\ref{fig:CE_comparison_516_560}.

A further scenario is that the hypothetical $6^-$ state is located at 75 keV above the 1999.0-keV state and the highly-converted $E1$ transition is thus obscured by the intense 75.3-keV $2^+_{1}\rightarrow 0^+_{g.s.}$ transition. If this were the case, we would observe excess intensity in the 75.3-keV $\gamma$-ray transition gated on transitions that depopulate the 1999.0-keV state when compared with spectra gated on transitions that run parallel to the 1999.0-keV state. We did not observe any excess $\gamma$-ray intensity in the 75.3-keV transition in these spectra, which indicates that the hypothetical $6^-$ state is not located 75 keV above the 1999.0-keV state.

We thus have found no evidence supporting the scenario of a hypothetical $6^-$ state located just above the 1999.0-keV state in $^{160}$Gd that is heavily populated by the $\beta$-decay of a negative-parity $^{160}$Eu parent.

We now discuss the option that the parent $^{160}$Eu has positive parity and undergoes a direct $\beta$-decay to the 1999.0-keV $(6^+)$ state in $^{160}$Gd. In this case, the log($ft$) for the $\beta$-decay to the 1999.0-keV state is log($ft$)=5.04(1) which is an allowed $\beta$-decay from a $J^\pi$= $5^+$ or $6^+$ parent state. Two possible configurations can reasonably be considered for the positive-parity parent: the $\pi5/2[532] \otimes \nu5/2[523]$ and $\pi5/2[413] \otimes \nu5/2[642]$ quasiparticle configurations, both of which can couple to create $K^\pi$=$5^+$ states. The $\pi5/2[532] \otimes \nu5/2[523]$ configuration involves proton and neutron orbitals that appear in the vicinity of the Fermi surface in $^{160}$Eu. For example, in $^{159}$Eu, the $5/2^+$ ground-state is built on the $\pi5/2[413]$ orbital, while a $5/2^-$ state built on the $\pi5/2[532]$ orbital lies only 190 keV above the ground-state \cite{nuclear_data_sheets_A_159}.

Theoretical predictions for the band-head energies and Gallagher-Moszkowski splittings~\cite{GM_rules} of relevant quasiparticle configurations calculated using a quasiparticle-plus-phonon model have been summarized in Ref.~\cite{Jain_review}. 
For $^{160}$Eu the calculations predict the $\pi5/2[413] \otimes \nu5/2[523]$  quasiparticle configuration triplet and singlet states to be the ground-state and at 17 keV excitation energy, respectively. This is consistent with the systematic trends and the assignments made by Hartley $et\;al.$~\cite{Hartley_PRL,Hartley_PRC}. No calculations were included in Ref.~\cite{Jain_review} for the $\pi5/2[532] \otimes \nu5/2[523]$ configuration in $^{160}$Eu.  It is, however, noteworthy that in the lighter-mass europium isotopes ($A=152,154,156$) the calculations show the $\pi5/2[413] \otimes \nu5/2[523]$ configuration (among others) varying in energy on the order of 200 keV between isotopes. In $^{152}$Eu, the $\pi5/2[532] \otimes \nu5/2[523]$ is actually predicted to be lower than the $\pi5/2[413] \otimes \nu5/2[523]$ configuration, although the structure of $^{152}$Eu, a transitional nucleus, is sufficiently different from the well deformed $^{160}$Eu to draw any firm conclusions from this.

However, given that the experimental data suggests a positive-parity $\beta$-decaying state in $^{160}$Eu, and despite the fact that the $\pi5/2[413]$ orbital is firmly assigned as the ground-state in the adjacent $^{159}$Eu, one may consider that the $\pi5/2[532] \otimes \nu5/2[523]$ configuration may lie at a similar or lower energy than the $\pi5/2[413] \otimes \nu5/2[523]$ configuration in $^{160}$Eu.
 If that were the case, the $\beta$-decay from the $\pi5/2[532] \otimes \nu5/2[523]$ configuration would be a $\nu5/2[523]\rightarrow \pi7/2[523]$ allowed transition to the previously-suggested $\pi^2(5/2[532],7/2[523])$ configuration in $^{160}$Gd, forming the 1999.0-keV $6^+$ state. Ref.~\cite{bohr_mottelson_ii} gives a prediction of log($ft$)=4.8 for the $\nu5/2[523]\rightarrow \pi7/2[523]$ transition, which matches well with the observed log($ft$)=5.04(1). Nearby, the $^{159}$Sm $\beta$-decay into $^{159}$Eu is understood to be a $\nu5/2[523]\rightarrow \pi7/2[523]$ allowed-unhindered transition with a log($ft$)=5.0 \cite{nuclear_data_sheets_A_159}, which matches the observed log($ft$) value for the $^{160}$Eu $\beta$-decay.

We also investigated if the available information on the low-spin $\beta$-decaying state in $^{160}$Eu might provide further insights on the question on the involved configurations. 
The spin-triplet coupling of the $\pi5/2[532] \otimes \nu5/2[523]$ configuration would lead to a lower-lying $J^\pi=0^+$ configuration which could be a potential candidate for the low-spin $\beta$-decaying state. The $\beta$-decay pattern suggests a $J=(1)$ assignment given the decays to states of spin $J=1$ and $J=2$, which would indicate the presence of a Newby shift \cite{newby_shift}. The spin-triplet coupling of the $\pi5/2[532] \otimes \nu5/2[523]$ configuration would consequently suggest a $J^\pi=1^+$ assignment for the low-spin isomer in $^{160}$Eu. Ref.~\cite{Hartley_PRC} suggested a $J^\pi=0^-$ or $1^-$ (depending on the presence of a Newby shift) configuration given the strong $\beta$-decay to the $(1^-)$ 2464.8 keV state. We also observe a strong decay to this state with log($ft$)=5.29(1) indicating an allowed decay.

The Alaga rules~\cite{Alaga_intensity_rules} can be used to estimate the $\beta$-feeding intensity to the $^{160}$Gd ground-state and attempt to constrain the spin and parity of the low-spin $^{160}$Eu parent. For an initial $^{160}$Eu configuration of $I_i=1$, $K_i=0$, the Alaga rules yield a $\beta$-feeding intensity of approximately 22\% to the ground-state of $^{160}$Gd. With this, the dominant $\beta$-decay to the 2464.8-keV ($1^-$) state has log($ft$)$\approx$5.5. With instead $I_i=1$, $K_i=1$, a $\sim$52\% $\beta$-intensity to the ground-state is expected, and the log($ft$) to the 2464.8-keV state then becomes log($ft$)$\approx$5.7. In both cases, when taking into account the ground-state feeding, the log($ft$) values indicate that the predominant decay to the 2464.8-keV state is an allowed transition with no parity change. However, since the ($1^-$) assignment of the 2464.8-keV state from Ref.~\cite{Hartley_PRC}  relied on the proposed $1^-$ assignment of the parent $^{160}$Eu, it is difficult to determine the parent state's parity with certainty using these considerations.


Our log($ft$) values support a $J=1$ assignment for the low-spin isomer of $^{160}$Eu, but it is difficult to determine the parity of the state. The $\beta$-decay pattern and observed log($ft$) values are consistent with both a positive and negative parity assignment for the $^{160}$Eu parent. 
While a $J^\pi$=$1^+$ assignment resulting from the unfavored coupling of the aforementioned $\pi5/2[532] \otimes \nu5/2[523]$ configuration with a Newby shift appears the favored solution given the experimental data, we cannot rule out a $J^\pi$=$1^-$ assignment resulting from some other unknown nearby configuration.


Based on all the available information, no firm conclusions on the configuration of the $\beta$-decaying states in $^{160}$Eu can be given. The naturally expected negative-parity configuration is in tension with the strong population of the positive parity-state at 1999.0 keV in $^{160}$Gd, which appears to be populated directly by an allowed transition. At the same time, while a positive parity configuration in $^{160}$Eu  would naturally decay to the 1999.0-keV state, it is not expected to be the lowest configuration based on the established systematics in this region. Further experimental and theoretical studies are needed to resolve this issue.


\begin{figure*}[!tbp!]
\includegraphics[width=1\textwidth]{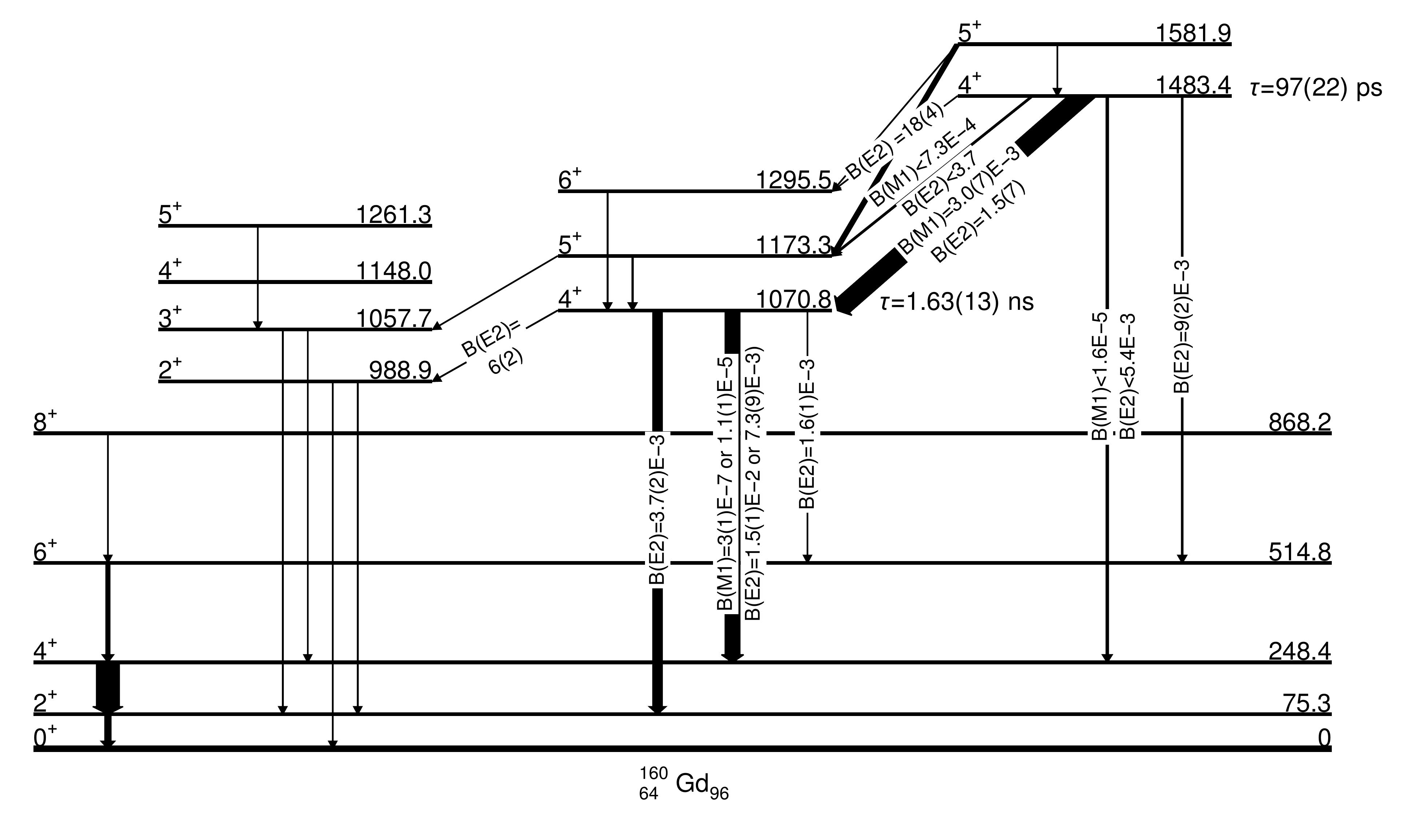}
\caption{\label{fig:simplified_levels}Simplified level scheme showing the levels and decays associated with the $K^\pi$=$4^+$ and $\gamma$-bands in $^{160}$Gd. Line widths indicate $\gamma$-ray intensities. B($M1$) and B($E2$) values are in units of W.u. Note that for transitions where the reduced transition rate has asymmetric error bars, the larger of the uncertainties is used here; Table \ref{tab:red_rates_sum} contains detailed reduced transition rates for these transitions.} 
\end{figure*}

\subsection{$K^\pi$=$4^+$ Band-heads}

The $4^+$ state at 1070.8 keV (the ``$4^+_I$" state) and the $4^+$ state at 1483.4 keV (the ``$4^+_{II}$" state) are both band-heads of $K^\pi$=$4^+$ bands. The band built on the $4^+_I$ state has been recently extended up to angular momentum 18 $\hbar$ \cite{Hartley_gd_coulex}. 
While the $K^\pi$=$4^+$ bands in even-even rare-earth nuclei have been originally interpreted as double-$\gamma$ phonons, it was pointed out by Burke \cite{Burke} that the experimental evidence from transfer reaction studies and transition rates is more consistent with these band-heads being hexadecapole phonons.

As shown in Ref.~\cite{Hartley_PRC}, the $K^\pi$=$4^+$ states in $^{160}$Gd seemingly break the isotopic trend of energies for the $K^\pi$=$4^+$ states that are candidates for hexadecapole vibrational states. In that context, Hartley $et\;al.$ ~\cite{Hartley_PRC} suggested that the two $K^\pi$=$4^+$ states in $^{160}$Gd result from the strong mixing of the lowest two $K^\pi$=$4^+$ two-quasiparticle configurations, $\nu^2(5/2[523],3/2[521])$ and $\pi^2(5/2[413],3/2[411])$. 

In the present work, we were able to establish several additional pieces of information that can help shed more light on the nature of the $K^\pi$=$4^+$ band-heads in $^{160}$Gd. Figure \ref{fig:simplified_levels} shows the part of the $^{160}$Gd level scheme relevant for this discussion. It shows the low-spin parts of the ground-state band, the $\gamma$-vibrational band, and the two $K^\pi$=$4^+$ bands. The newly measured lifetimes of the $4^+_I$ and $4^+_{II}$ states are indicated as well as the transition strengths in W.u. for the transitions depopulating the two $4^+$ states. For the 822.2-keV $4^+_{I} \rightarrow 4^+_{gsb}$ transition, B($M1$) and B($E2$) strength are given for the two possible solutions of the mixing ratios. The B($M1$) strength for the $4^+_{II} \rightarrow 4^+_I$ transition is $3.0(7)\times10^{-3}$ W.u. = $5(1)\times10^{-3}$ $\mu^2_N$. For the 309.9-keV $4^+_{II} \rightarrow 5^+_{I}$ and  1235.0-keV $4^+_{II} \rightarrow 4^+_{gsb}$  transitions upper limits for the B($M1$) and B($E2$) strength are established using the assumption of pure $M1$ or $E2$ character, respectively.

The most striking result is that the $4^+_{II}$ state decays with collective $E2$ transitions to the $4^+_{I}$ and $6^+_{I}$ states and that there is a weak B($M1$) transition from the $4^+_{II}$ state to the $4^+_{I}$ state. At the same time, both $4^+_I$ and $4^+_{II}$ states decay with non-collective $E2$ transitions to the ground-state band. We also observe the decay from the lower $K^\pi$=$4^+$ band to the $\gamma$-band with the transition from the $4^+_I$ to the $2^+_\gamma$ state weakly collective in nature with a B($E2$) strength on the order of 1 W.u. 

The newly measured enhanced B($E2$) value between the two $4^+$ states are consistent with a strong overlap in wavefunctions of the two $4^+$ states. However, a strong two-level mixing scenario for just the band-heads or the first few states of the rotational bands seems to be at odds with the interpretation that the rotational band built on the $4^+_I$ state is dominated by the $\nu^2(5/2[523],3/2[521])$ two-quasineutron configuration above spin 6 $\hbar$  \cite{Hartley_gd_coulex},  based on the observed small B($M1$)/B($E2$) ratios. 
If only the $4^+$ band-heads were to be mixed strongly with an interaction of $V\approx 200$  keV, the rotational structure of the band would be changed significantly near the band-head due to the large energy shift of the $4^+$ state. 
At a minimum, a strong expansion of the rotational sequence would be observed for the lower band and a strong compression for the higher band. This is incompatible with the observed energy spectrum. Thus the bands would have to mix strongly over a large spin range so that the bands are shifted as a whole, or the mixing has to be weak. However, a strong mixing scenario of almost degenerate quasiparticle configurations $\nu^2(5/2[523],3/2[521])$ and $\pi^2(5/2[413],3/2[411])$ over a large spin range seems to be inconsistent with the interpretation put forward in Ref. \cite{Hartley_gd_coulex} based on the alignment and B($M1$)/B($E2$) ratios.

Weak mixing of the $K^\pi$=$4^+$ bands would only lead to a few percent admixture of the higher band wave function into the lower band. In order to reproduce the strong B($E2$) transition of $\sim$10 W.u. from the $4^+_{II}$ state to the $4^+_I$ and $6^+_I$ states, the deformation difference between the bands would have to be of order $\Delta\beta\approx 25 \%$, which appears to be inconsistent with the deformations expected for both two-quasiparticle configurations based on the Nilsson diagram as well as the fact that the moment of inertia of both bands seem almost identical near the band-head, based on the energy differences of the $5^+$ and $4^+$ band members in each band.

While the simple two-level-mixing descriptions for the band-heads are at odds with the experimental data, quasiparticle-phonon nuclear model (QPNM) calculations actually provide a rather satisfactory description of the data for the band-heads. QPNM calculation in Ref.~\cite{soloviev97} predict the energies of the lowest two $K^\pi$=$4^+$ hexadecapole vibrational states (with $>90\%$ purity) in $^{156}$Gd (1.5 MeV, 1.9 MeV),  $^{158}$Gd (1.4 MeV, 1.9 MeV), and $^{160}$Gd (1.1 MeV, 1.5 MeV), reproducing the experimentally observed trend quite accurately. 
The two $K^\pi$=$4^+$ states in $^{160}$Gd at 1070.8 keV and 1483.4 keV are thus not breaking a trend as suggested in Ref. \cite{Hartley_PRC} but follow the theoretical predictions well, which take into account the evolution of the underlying single-particle structure in the isotopic chain. In all of these $4^+$ states, the $\pi^2(5/2[413],3/2[411])$ and $\nu^2(5/2[523],3/2[521])$ configurations are dominant, although with changing relative contributions.

The $\pi^2(5/2[413],3/2[411])$ configuration dominates the lower $4^+$ wavefunction in $^{156}$Gd and $^{158}$Gd (supported by ($\alpha$,$2n$) \cite{KONIJN_156Gd} and ($t$,$\alpha$) \cite{burke_158Gd} data) with $> 80\%$, the higher-lying $4^+$ state is comparably dominated by the $\nu^2(5/2[523],3/2[521])$ configuration (supported by ($d,t$) (unpublished data cited in Ref.~\cite{BACKLIN1982189}) and ($d$,$p$) \cite{GREENWOOD1978327} data). In $^{160}$Gd, the situation is different because both $4^+$ states have about equal admixture of proton and neutron quasiparticle configurations~\cite{soloviev97}.

Soloviev $et\;al.$ \cite{soloviev94} also calculated the transition strength for the decay of the two lowest $K^\pi$=$4^+$, $4^+$ states in $^{156,158}$Gd and found B$(M1)$ values from the higher lying $K^\pi$=$4^+$ state to the lower one of $0.04\;\mu_N^2$ in both nuclei, reproducing the experimental branching ratios for the decay of the $4^+_{II}$ states reasonably well. Unfortunately, there are no predictions available for $^{160}$Gd and since the wavefunctions of the hexadecapole vibrations are similar in $^{156,158}$Gd but significantly different than in $^{160}$Gd, it is difficult to make an extrapolation. However, Ref. ~\cite{soloviev97} remarks that $M1$ transitions are to be expected in the range $10^{-4}-10^{-1}\;\mu_N^2$. At the same time the intensity of the $M1$ transition is generally predicted to be dominating over the $E2$ intensity \cite{soloviev94}. In the case of the 412.6 keV transition between the two $K^\pi=4^+$, $4^+$ states, the $M1$ intensity is $6(3)$ times larger than the $E2$ intensity, in line with this prediction.

As pointed out by Burke~\cite{Burke}, the hexadecapole nature of these states is also indicated by Interacting Boson Model (IBM) calculations involving a g-boson~\cite{VanIsacker82}. These sdg-IBM calculations are also able to predict the energies of the low-lying $4^+$ states in the even-even Gd isotopes $^{154-158}$Gd.
The interpretation as hexadecapole phonons for the $4^+_{I,II}$ states in $^{160}$Gd is also consistent with the observed strong transition from the $4^+_I$ state to the $2^+_\gamma$ state \cite{Burke,devi91,VanIsacker82}. It should also be noted that inelastic scattering data of $^{160}$Gd \cite{ichihara_160Gd} required a strong hexadecapole component to explain the observed cross-section to the $4^+_\gamma$ state. 

Analysis of the available experimental data appears to support the $K^\pi$=$4^+$ band-heads as being hexadecapole in nature and is less consistent with the interpretation of these band-heads as being strongly-mixed quasiparticle configurations. However, further theoretical investigations of the $K^\pi$=$4^+$ bands in $^{160}$Gd are required to fully understand the extensive experimental data. Transfer data could also help to experimentally elucidate the underlying quasiparticle structure of the $4^+$ band-heads. The hexadecapole component for excitations into the $\gamma$-band mentioned in Ref.~\cite{ichihara_160Gd} makes further inelastic scattering data on $^{160}$Gd to measure the $E4$ strength to the $K^\pi=4^+$ of interest as well. 


\section{Conclusion}\label{Sec:Concl}

High-statistics $\gamma$-ray spectroscopy data following the $\beta$-decay of $^{160}$Eu was carried out with the GRIFFIN spectrometer at the TRIUMF-ISAC facility. Ten new transitions and two new levels in $^{160}$Gd associated with the $\beta$-decay of the high-spin $\beta$-decaying state in $^{160}$Eu were observed, including two transitions from the lowest $K^\pi$=$4^+$ band to the $\gamma$-band. 

The lifetimes of the two known $K^\pi$=$4^+$ band-heads in $^{160}$Gd were measured for the first time, and the majority of levels and transitions recently published were observed. Based on the lifetime limit of $\tau<100$ ps for the 1999.0 keV state in $^{160}$Gd 
as well as its decay pattern, we conclude that it is a $J^\pi$=$(6^+)$ state. However, this raises questions as to the nature of the $\beta$-decaying states in $^{160}$Eu, which could not be conclusively resolved. Angular correlation measurements allowed the determination of mixing ratios for several transitions, in particular constraining the B($M1$) and B($E2$) strength for the transition between the $K^\pi$=$4^+$ band-heads. 

Overall, the energies and transition strengths between the $K^\pi$=$4^+$ bands and the observed transition from the lower $4^+$ band-head to the $\gamma$-band are consistent with the two $K^\pi$=$4^+$ band-heads being interpreted as hexadecapole phonons, as predicted in the Quasiparticle Phonon Model. However, more detailed experimental and theoretical investigations of the structure of the band-head and the rotational bands built on them will be required to further scrutinize this interpretation. At the same time, further experimental data, such as transfer cross-sections and transition rates from inelastic scattering data, are needed to clarify the underlying quasiparticle configurations in these states.


\section*{Acknowledgements}
 The authors would like to thank the operators at the TRIUMF-ISAC facility for providing the radioactive beam. R.K. would like to thank R.M. Clark and A.O. Macchiavelli for insightful discussions. This work was supported by the Natural Sciences and Engineering Research Council of Canada (NSERC), the Canada Research Chairs Program, NSERC Discovery Grants No. SAP-IN-2014-00028, SAP-IN-2017-00039, and RGPAS 462257-2014, and the NSERC CREATE Program IsoSiM (Isotopes for Science and Medicine). 
 The GRIFFIN spectrometer was funded by the Canada Foundation for Innovation (CFI), TRIUMF, and the University of Guelph. TRIUMF receives federal funding via a contribution agreement with the National Research Council of Canada (NRC). This work was supported by the U.S. Department of Energy (DOE) under contract no. DE-FG02-93ER40789. The Tennessee Tech. University team was supported by the Office of Nuclear Physics, U.S. DOE under contract DE‐SC0016988. P.H.R. was supported by the UK STFC via Grants No. ST/L005743/1 and No. ST/P005314. P.H.R. also acknowledges support from the UK Government Department of Business, Energy, and Industrial Strategy via the National Measurement System. Figures \ref{fig:level_scheme_high}, \ref{fig:level_scheme_low}, and \ref{fig:simplified_levels} were created using the SciDraw package \cite{SciDraw}.


\section*{Appendix}

Reduced transition rates derived from the measured lifetimes and mixing ratios measured in this work are given in Table~\ref{tab:red_rates_sum}. Conversion electron coefficients were applied when the multipolarity of a transition is known; for transitions where the multipolarity is not known, the lowest multipolarity conversion electron coefficient was applied. Upper limits are given for transitions where no mixing ratio was measured for the 1070.8, 1483.4, and 1581.9 keV states. For the 1999.0 keV state, lower limits for the reduced transition rates are given based on the measured lifetime of $\tau<100$ ps and assumed pure $M1$/$E2$ transitions when no mixing ratio was measured. The 515.6 keV transition values were calculated using $\delta=0$.

\begin{longtable*}[!t!]{c|c|c|c|c|c|c|c}
     \caption{Reduced transition rates for transitions from the 1070.8, 1483.4, 1581.9, and 1999.0 keV states based on measured lifetimes and mixing ratios.}\\
    \hline \hline
    Level & Lifetime & Transition& $\delta$ & B($M1$)  & B($E2$)  & B($M1$) & B($E2$)   \\ 
    (keV) & (ps) & (keV) & & ($\mu_N^2$) & ($e^2\mathrm{fm}^4$) & (W.u.) & (W.u.) \\
    \hline 
    \endfirsthead
    \caption{$(Continued.)$}\\
    \hline \hline
    Level & Lifetime & Transition& $\delta$ & B($M1$)  & B($E2$)  & B($M1$) & B($E2$)   \\ 
    (keV) & (ps) & (keV) & & ($\mu_N^2$) & ($e^2\mathrm{fm}^4$) & (W.u.) & (W.u.) \\
    \hline 
    \endhead
    \endfoot
    \hline\hline
    \endlastfoot
    1070.8 & 1630(130) &81.5 & - & - & $330(110)$ & - & $6(2)$ \\
    
    && 555.7 & - & - & $8.1(5)\times10^{-2} $ & - & $1.6(1)\times10^{-3}$ \\
    
     && 822.2 & $-8.1^{+2}_{-183}$ & $5.7^{+5}_{-25}\times10^{-7}$ & $0.79^{+5}_{-7}$ & $3.2^{+3}_{-14}\times10^{-7}$ & $1.5(1)\times10^{-2}$\\ 
         
    && 822.2 & $0.94^{+9}_{-8}$ & $2.0(2)\times10^{-5}$ & $0.38(4)$ & $1.1(1)\times10^{-5}$ & $7.3(9)\times10^{-3}$ \\ 
    
    && 995.5 & - & - & $0.19(1)$ & - & $3.7(2)\times10^{-3}$ \\
         \hline

         1483.4 & 97(22) &  187.6 & - & - & $950(230)$ & - & $18(4)$ \\
         
         && 309.9 & - & $<1.3\times10^{-3}$ & $<195$ & $<7.3\times10^{-4}$ & $<3.7$ \\
         
          && 412.6 & $0.41(10)$ & $5(1)\times10^{-3}$ & $76^{+36}_{-20}$  & $3.0(7)\times10^{-3}$  & $1.5^{+7}_{-4}$ \\

         
         && 968.8 & - & - & $0.5(1)$ & - & $9(2)\times10^{-3}$ \\
         
         && 1235.0 & - & $<3.0\times10^{-5}$ & $<0.28$ & $<1.6\times10^{-5}$ & $<5.4\times10^{-3}$ \\
         
        \hline
    
        1581.9 & 160(20) & 98.6 & $|\delta|<0.41$ & $>5.4\times10^{-2}$ & $<2.0\times10^{4}$ & $>3.0\times10^{-2}$ & $<300$ \\
        
        && 286.2 & - & $<4.1\times10^{-3}$ & $<7.2\times10^{3}$ & $<2.2\times10^{-3}$ & $<13.9$ \\
        
        && 408.5 & - & $<7.1\times10^{-4}$ & $<62$ & $<4.0\times10^{-4}$ & $<1.2$\\

      
      
         
         
         \hline
         
         1999.0 & $<100$ & 300.2 & - & $>1.0\times10^{-4}$ & $>16$ & $>5.6\times10^{-5}$ & $>0.32$ \\
         
         && 417.0 & - & $>1.1\times10^{-3}$ & $>83$ & $>5.7\times10^{-4}$ & $>1.7$ \\
         
         && 450.4 & - & $>1.2\times10^{-4}$ & $>8$ & $>6.4\times10^{-5}$ & $>0.14$ \\
         
          && 515.6 & $|\delta|<3.6\times10^{-4}$ & -  & $>140$ & - & $>2.7$ \\
         
         && 605.7 & - & - & $>1.0$ & - & $>2.0\times10^{-2}$ \\
         
         && 737.8 & - & $>1.3\times10^{-4}$ & $>3.5$ & $>7.3\times10^{-5}$ & $>6.8\times10^{-2}$ \\
         
        & & 825.7 & - & $>1.4\times10^{-5}$ & $>0.3$ & $>7.9\times10^{-6}$ & $>5.7\times10^{-3}$ \\
         
         && 1484.0 & - & $>3.0\times10^{-7}$ & $>1.9\times10^{-3}$ & $>1.6\times10^{-7}$ & $>3.8\times10^{-5}$ 
         
    \label{tab:red_rates_sum}
    
\end{longtable*}



\bibliography{160Eu-GRIFFIN.bib} 

\end{document}